\documentclass[hyper,letterpaper]{JHEP3} 

\JHEPspecialurl{http://jhep.sissa.it/JOURNAL/JHEP3.tar.gz}

\newcommand{\epp}{\ensuremath{e^{\pm}}}

\newcommand{\gsim}{ \mathop{}_{\textstyle \sim}^{\textstyle >} }
\newcommand{\lsim}{ \mathop{}_{\textstyle \sim}^{\textstyle <}}
\newcommand{\vev}[1]{ \left\langle {#1} \right\rangle }

\newcommand{\be}{\begin{eqnarray}}
\newcommand{\ee}{\end{eqnarray}}
\newcommand{\tev}{\rm \, TeV}
\newcommand{\gev}{\rm \, GeV}
\newcommand{\mev}{\rm \, MeV}
\newcommand{\kev}{\rm \, keV}

\usepackage{epsfig}

\newcommand\fverb{\setbox\pippobox=\hbox\bgroup\verb}
\newcommand\fverbdo{\egroup\medskip\noindent%
            \fbox{\unhbox\pippobox}\ }
\newcommand\fverbit{\egroup\item[\fbox{\unhbox\pippobox}]}

\newbox\pippobox

\title{MiXDM: Cosmic Ray Signals from Multiple States of Dark Matter}

\author{Ilias Cholis and Neal Weiner\\
     Center for Cosmology and Particle Physics, 
  Dept. of Physics, New York University, \\
New York, NY 10003\\
    E-mail: \email{nw32@nyu.edu, ijc219@nyu.edu}}

\abstract{Recent data from cosmic ray experiments such as PAMELA, Fermi, ATIC and PPB-BETS all suggest the need for a new primary source of electrons and positrons at high ($\gsim 100 \gev$) energies. Many proposals have been put forth to explain these data, usually relying on a single particle to annihilate or decay to produce \epp. In this paper, we consider models with multiple species of WIMPs with significantly different masses. We show if such dark matter candidates $\chi_i$ annihilate into light bosons, they naturally produce equal annihilation rates, even as the available numbers of pairs for annihilation $n_{\chi_i}^2$ differ by orders of magnitude. We argue that a consequence of these models can be to add additional signal naturally at lower ($\sim 100 \gev$) versus higher ($\sim \tev$) energies, changing the expected spectrum and even adding bumps at lower energies, which may alleviate some of the tension in the required annihilation rates between PAMELA and Fermi. These spectral changes may yield observable consequences in the microwave Haze signal observed at the upcoming Planck satellite. Such a model can connect to other observable signals such as DAMA and INTEGRAL by having the lighter (heavier) state be a pseudo-Dirac fermion with splitting 100 keV (1 MeV). We show that variations in the halo velocity dispersion can alleviate constraints from final state radiation in the galactic center and galactic ridge. If the lighter WIMP has a large self-interaction cross section, the light-WIMP halo might collapse, dramatically altering expectations for direct and indirect detection signatures.}

\begin{document}



\section{Introduction}\label{sec:intro}
Signals from a wide range of sources have prompted a rethinking of the nature of dark matter. The positron fraction of cosmic rays has shown a steep rise \cite{Adriani:2008zr}, implying the presence of a new primary source of antimatter. The Fermi \cite{Abdo:2009zk}, ATIC \cite{:2008zzr} and PPB-BETS \cite{Torii:2008xu} experiments all show a harder spectrum of \epp at high ($\gsim 100 \gev$) energies than had been predicted from secondary production, with a break at $\sim$ 1 TeV seen at Fermi and HESS  \cite{Collaboration:2008aaa} \cite{Aharonian:2009ah}. Data from INTEGRAL suggest a new source of low energy positrons, while the DAMA modulation signal has extended to 8.2$\sigma$.

Remarkably, there is a simple ``unified'' framework that explains all of these anomalies \cite{ArkaniHamed:2008qn}. One of the principle changes is that rather than annihilating to standard model channels,  the WIMP can annihilate dominantly into a new light, unstable boson, which in turn maintains equilibrium with the standard model and decays through a small mixing parameter \cite{Finkbeiner:2007kk}. The light mass kinematically prevents production of antiprotons \cite{Cholis:2008vb}, allowing consistency with the PAMELA antiproton measurements \cite{Adriani:2008zq}. Such scenarios have been widely discussed in the context of present dark matter anomalies \cite{Finkbeiner:2007kk,Pospelov:2007mp,Cholis:2008vb,ArkaniHamed:2008qn,Pospelov:2008jd,Nelson:2008hj,Cholis:2008qq,Nomura:2008ru}. 

In SUSY models, the LSP is stable by R-parity, and thus there is a unique WIMP candidate. In contrast, in these models, the symmetry stabilizing the dark matter against decay is not R-parity, and thus there is no compelling reason that there should be a single stable neutral particle. Indeed, it is very possible - perhaps even likely - that there are a variety of parities in the theory, and that many different particles together comprise the dark matter. We shall see that in models where dark matter freezes out by annihilation into a light boson, the annihilation rates of different components of dark matter {\em are naturally equal}, even though $n_\chi^2$ can differ between the species by orders of magnitude. As a consequence, the spectra of positrons and electrons can have interesting structures, including crests and troughs. This phenomenon arises naturally, without any tuning of the parameters. Such a model can alleviate some tension between PAMELA, which seems to require somewhat larger cross sections at lower energy, and Fermi, which requires lower cross sections but extending to higher energy.

In such models, one can explain the INTEGRAL signal with a heavy ($\sim \rm TeV$) mass state exciting into an MeV excited state as in \cite{Finkbeiner:2007kk}, while explaining DAMA through a lighter ($\sim 100 \gev$) state exciting into a $\sim 100 \kev$ excited state, as in \cite{TuckerSmith:2001hy,Chang:2008gd}. This scenario, to employ multiple states for inelastic and exciting dark matter (which we refer to as MiXDM) was proposed in \cite{Katz:2009qq}. Although DAMA and INTEGRAL both play roles in motivating the spectrum considered, we shall not focus on these aspects, focusing instead on the higher energy cosmic ray signals (and their synchrotron counterparts). Because of the motivation in \cite{Katz:2009qq},  we shall generically refer to these models, in which multiple states annihilate comparably into light bosons, as ``MiXDM'' models.

In this paper, we shall explore the possibility of testing such a scenario. In section \ref{sec:multicompmodel}, we will review the setup and in section \ref{sec:mixmodels} the model building issues, including models where different annihilation channels arise. In section \ref{sec:multicompcr}, we discuss the cosmic ray signatures from such a scenario, and discuss the interesting spectral features that can arise. In \ref{sec:Haze}, we go beyond the $e^+e^-$ signals, we shall argue that signatures in the microwave haze of the inner galaxy may help to show if the dark sector has multiple components. In section \ref{sec:3_masses} we consider further generalizations, with more states, while in section \ref{sec:scatter}, we note the important changes to direct detection experiments that can arise. Finally, we conclude in section \ref{sec:Conclusions}.

\section{Dark Matter with Multiple Components}
\label{sec:multicompmodel}
The idea that dark matter has multiple components is not new. In \cite{Duda:2001ae,Duda:2002hf}, it was argued that a subdominant component of dark matter in the halo could still give significant signals at direct detection experiments, and indirect signals related to WIMP capture (such as in the Earth). \cite{Zurek:2008qg} argued that a sector combining a TeV WIMP together with a GeV mass WIMP could explain PAMELA and DAMA. As we have already noted, \cite{Katz:2009qq} constructed a model in which a TeV particle was responsible for the high energy \epp signatures, as well as the 511 keV emission observed by INTEGRAL, while a separate particle with a $100 \kev$ splitting would explain DAMA. 

Our effort here is not to focus on the direct detection signals of multiple states, but rather their {\em indirect} signals. We shall see that if dark matter freezes out and annihilates presently into a light force carrier, the annihilation rates are naturally equal. Because the lighter WIMP has its signal compressed into a narrower energy range, this can lead to an enhancement of \epp  at lower energies, possibly even yielding a bump in the $e^+$ or $e^+ + e^-$ spectrum.

\subsection{Annihilation rates}\label{sec:annih_rates}

\begin{figure}[h]
\centering
\begin{tabular}{cc}
\epsfig{figure=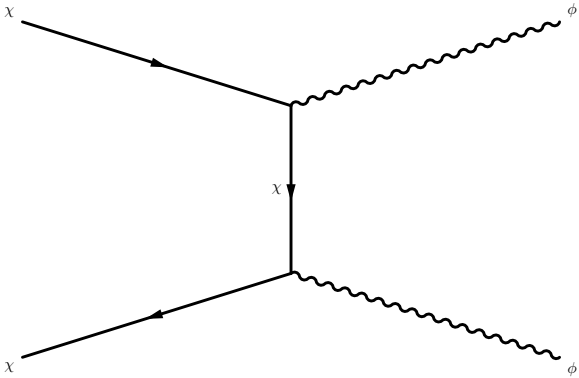,width=3.0in} &
\end{tabular}
\caption{Perturbative annihilation of dark matter into $\phi$.}
\label{fig:freezeout}
\end{figure}

Let us begin by sketching out the process described in \cite{Finkbeiner:2007kk} whereby dark matter freezes out through annihilations into a new metastable force carrier $\phi$, to which it couples with strength $g$. The annihilation proceeds through the diagram shown in figure \ref{fig:freezeout}, and  has a cross section (at high energy) that scales as 
\be
\sigma v \sim \frac{\alpha^2}{M_{\chi}^2}.
\ee
The dark matter stays in thermal equilibrium with the standard model through the interactions of $\phi$ with the standard model, in particular the process $e \phi \leftrightarrow e \gamma$ or $\phi \phi \leftrightarrow f \bar f$, depending on the model. As was emphasized in \cite{Finkbeiner:2007kk,Pospelov:2007mp,Finkbeiner:2008gw,ArkaniHamed:2008qp}, this allows the ``WIMP miracle'' to function, even though the WIMP has no strong interactions to the standard model. Thus, one has the usual relation \cite{kolbandturner} that the dark matter density scales inversely with cross section. I.e.,
\be
\rho \sim \frac{1}{\vev{\sigma v}} \sim M_\chi^2,
\ee
where we focus only on the dominant mass scaling. The annihilation rate of the dark matter in the present universe then simply scales as
\be
n_\chi^2 \vev{\sigma v} \sim \left(\frac{\rho}{M_\chi}\right)^2 \frac{1}{M_\chi^2} \sim {\rm constant}.
\ee
Remarkably, because the number density scales as mass, the leading dependencies cancel precisely. Thus one sees immediately that {\em in scenarios where the dark matter freezes out through annihilations into a light force carrier, the annihilation rates of species of different mass are naturally equal.} \footnote{Note that while we are focused on models with new force carriers, this would also be true for annihilations into W's, Z's, or even light fermions, so long as the WIMP mass were suitably large compared to the particle into which it annihilated.}

There are of course caveats to this. We are assuming that the coupling $g$ is the same between the states. This arises automatically if $\phi$ is a boson of a new gauge group under which $\chi_{i}$ have the same representation, or if there is a symmetry relating the different $\chi_i$, broken softly by their masses. The Sommerfeld effect \cite{sommerfeld,Hisano:2004ds,Cirelli:2008id,MarchRussell:2008yu,ArkaniHamed:2008qp,Pospelov:2008jd}, which is often invoked to explain the high cross sections needed to explain the cosmic ray data, can for instance have already saturated for one WIMP and not the other, or be in a $v^{-2}$ resonance region for one and not the other. Factors of two can arise as well for fermions versus scalars, or if the representations of the groups are not identical. Finally, there is the usual logarithmic dependence on mass. That said, it is important to note that the {\em starting point} for consideration is that the annihilation rates should be comparable, and such effects can be invoked to move away from that.

\section{Models of MiXDM}\label{sec:mixmodels}
Scenarios where multiple states are present are trivial to construct. Probably the easiest means to do this is to WIMPs which are charged under a new $U(1)_{dark}$, with gauge field $\phi_\mu$, as was first considered by  \cite{Holdom:1985ag}. To this we add a small $\lsim \rm GeV$ mass to the gauge boson, and a small splitting $\delta$ between the Majorana states for the WIMPs. 
\be
{\cal L} = \bar \chi_i \not \! \! D \chi_i + \frac{1}{4}F^d_{\mu\nu} F^{d\mu\nu}+\epsilon F_{\mu\nu} F^{d\mu\nu}+m^2 \phi_{ \mu} \phi^{\mu}+M_i \bar \chi_i \chi_i+\delta_i \chi_i \chi_i.
\label{eq:lagrangian}
\ee
One can also trivially replace the fermionic WIMP with a complex scalar.
\be
{\cal L} = ( D_\mu \chi_i )^* D_\mu \chi_i + \frac{1}{4}F^d_{\mu\nu} F^{d\mu\nu}+\epsilon F_{\mu\nu} F^{d\mu\nu}+m^2 \phi_{ \mu} \phi^{\mu}+M^2_i  \chi_i^* \chi_i+M_i \delta_i \chi_i \chi_i .
\label{eq:lagrangianscalar}
\ee
Here $i$ indexes the WIMP species, and we have suppressed the necessary dark Higgs mechanism. Reduced to a single species with $\delta \approx 1 {\rm MeV}$, this is just the original XDM model proposed by  \cite{Finkbeiner:2007kk}. To explain INTEGRAL, such a model requires a double excitation. If the Milky Way halo has a uniform and low velocity dispersion, it is difficult to achieve this, as pointed out by \cite{Pospelov:2007xh}, prompting development of single excitation models \cite{Finkbeiner:2007kk,ArkaniHamed:2008qn} and inverted models \cite{Chen:2009ab,Finkbeiner:2009mi}, which more easily satisfy the INTEGRAL rate due to the naturally large scattering cross section from the light mediator \cite{Finkbeiner:2007kk}.  Some recent simulations involving baryons show a significant increase of the velocity dispersion in the galactic center \cite{fabioprivate,RomanoDiaz:2008wz,RomanoDiaz:2009yq,Abadi:2009ve,Pedrosa:2009rw}, which can alleviate these constraints \cite{Chen:2009av, robinprogress}, possibly opening the possibility of double-excitations, although this is unclear.

Meanwhile, the interaction with the standard model can allow us to construct models of inelastic dark matter \cite{ArkaniHamed:2008qn}. Since heavier WIMPs have lower modulation fraction, the constraints on them are stronger \cite{Chang:2008gd,MarchRussell:2008dy,Cui:2009xq,Cline:2009xd,Collaboration:2009xb}. While simulations and simulation-inspired halo models allow for heavier WIMPs \cite{MarchRussell:2008dy,kuhlen}, having two WIMPs is a simple approach to this problem as well. Note that the annihilation cross section goes as $\alpha^2_{dark}/M_i^2$ while the scattering cross section scales as $\epsilon^2/m_\phi^4$. Thus, even a subdominant WIMP can have significant scattering cross sections, even beyond the natural ``order one'' level considered by \cite{Duda:2001ae}.

Thus, taking the above Lagrangian with two species one arrives at the MiXDM scenario, which, suitably supersymmetrized, is just the setup proposed by \cite{Katz:2009qq}. 

We note that if there is an approximate symmetry protecting the WIMPs from both $M$ and $\delta$, it is natural that $\delta_i/M_i$ is constant, or at least roughly so.

An interesting generalization of this would be to a situation with {\em two} U(1)'s in the dark sector (or even more generally, each WIMP would have its own U(1)). I.e., we would modify \ref{eq:lagrangian} by
\be \frac{1}{4}F^d_{i \mu\nu} F_i^{d\mu\nu}+\epsilon F_{\mu\nu} F_i^{d\mu\nu}+m_i^2 \phi_{i \mu} \phi_i^{\mu}
\ee
In this case, it is reasonable to expect that the soft breaking of the masses for $\chi$ would radiatively feed into determining the masses of the $\phi$. Thus, one would expect $m_{i}/M_i = {\rm constant}$. This could be generated for instance when the WIMPs, themselves, serve as gauge mediators for the masses of the $\phi$'s as described by \cite{ArkaniHamed:2008qp,Morrissey:2009ur}, but not in cases where the mediation was dominated by the effective D-terms, as in \cite{Baumgart:2009tn,Katz:2009qq,Cheung:2009qd}.

Finally, we can consider situations where $\phi$ is a scalar, and there is a $Z_2$ symmetry between the two WIMPs. I.e., we could consider
\be
{\cal L} = \bar \chi_i \not \!   \partial \chi_i + \partial_\mu \phi^* \partial^\mu \phi +\lambda \phi^2 h^\dagger h+m^2 \phi^2+(g \phi+M_i) \bar \chi \chi+\delta_i \chi \chi 
\label{eq:lagrangianscalarscalar}
\ee
Which can be generalized to a multi-$\phi$ theory trivially, in which case again, one could argue that the $\phi_i$ masses would be proportional to the $\chi_i$ masses. The scalar mediated theories can yield XDM phenomenology, but is difficult to produce the iDM phenomenology naturally \cite{Finkbeiner:2008qu}. Since our focus is on cosmic rays, the particular Lorentz representation is not especially crucial, but a scenario with the DAMA signal likely fits best with vectors.

We should note that there are a wide variety of studies on the possibility of light force carriers. \cite{Boehm:2003hm} first studied them in the context of MeV scale dark matter, as an explanation of INTEGRAL. While XDM is limited to $\sim$ TeV masses for $\chi$ and $100 \mev \sim \gev$ for $\phi$, \cite{Pospelov:2007mp} considered the setup for a much wider kinematical range, including MeV WIMPs and $\sim 20 \gev$ force carriers. Decaying models have also invoked new light force carriers \cite{Ruderman:2009ta,Ruderman:2009tj} While we focus on cosmic ray signals here, a broad set of collider and other accelerator studies \cite{Pospelov:2008zw,Baumgart:2009tn,Batell:2009yf,Reece:2009un,Yin:2009mc,Cheung:2009su,Batell:2009jf} as well as astrophysical signatures \cite{Liu:2008ci,Batell:2009zp,Schuster:2009au,Schuster:2009fc,Meade:2009mu,Yin:2009yt} will be essential in testing this picture. (For a broad review of existing and future limits, see \cite{Schuster:2009au}.)

\section{Electron and Positron Signals of Multiple WIMPs} \label{sec:multicompcr}
\begin{figure}[h]
\centering
\begin{tabular}{cc}
\epsfig{figure=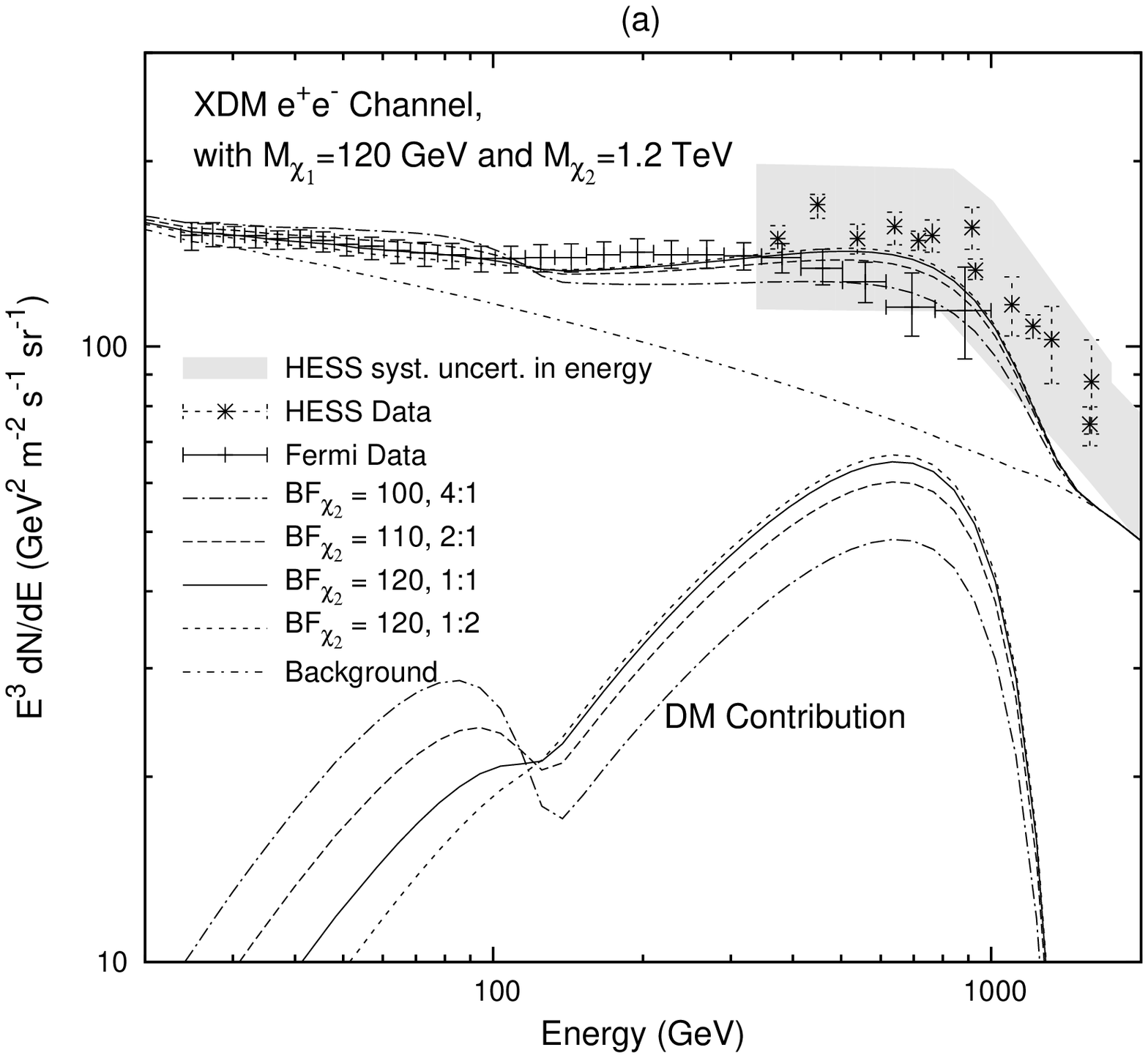,width=3.0in} &
\epsfig{figure=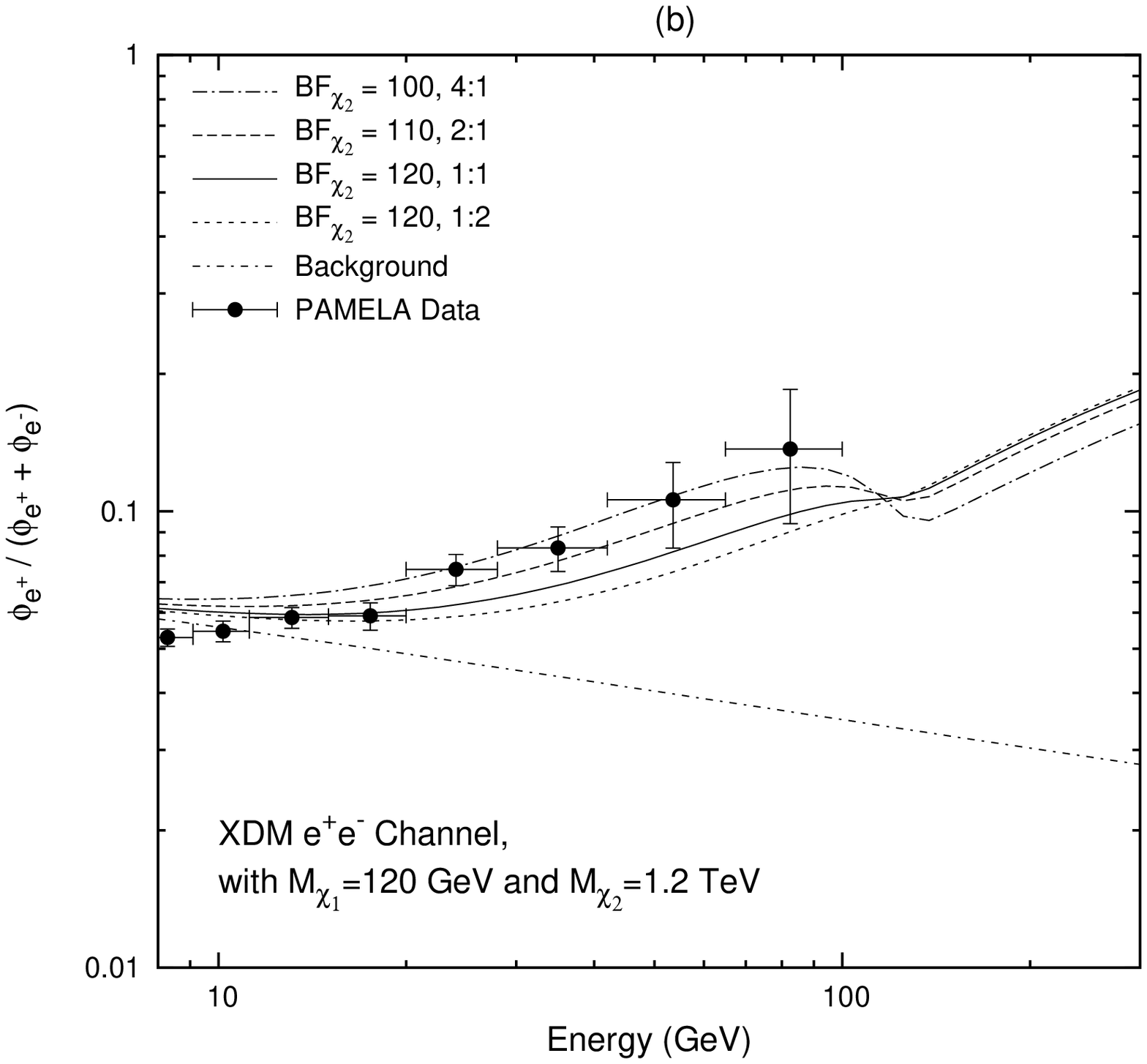,width=3.0in} 
\end{tabular}
\caption{Fermi/HESS electron plus positron ({\em left}) and PAMELA positron ({\rm right}) signals from annihilation of 2 XDM species with $M_{\chi_{1}}=120$ GeV and $M_{\chi_{2}}=1.2$ TeV through $e^{\pm}$ channel. BF specifies the cross section of the heavier WIMP while $x:y$ specifies the lighter WIMP has an annihilation rate $x/y$ times that of the heavier.}
\label{fig:WIMPspectra}
\end{figure}

Before progressing to consider in details the electron signals from multiple WIMPs, we can sketch out what sort of range of signals is possible. The simplest scenario is to consider two WIMPs, of mass $M_{1} \sim 100 \gev$ and $M_2 \sim 1 \tev$, annihilating through the channel $\chi \chi \rightarrow \phi \phi$ with $\phi \rightarrow e^+ e^-$ (which naturally occurs for $m_\phi < 2 m_\mu$). We calculate the cosmic ray spectra as described in appendix \ref{sec:CRprop}, and show the results in figure \ref{fig:WIMPspectra}. Here we see immediately the important consequence of this scenario. Although the standard case of a heavy WIMP annihilating through this channel yields very hard spectra, the inclusion of the lighter WIMP leads to a softening. Indeed, with a slightly higher rate than equal (which can arise, as described, from scalar/fermion differences, or differences in the Sommerfeld effect), bump like features can appear in the spectrum. This is possibly the most important point of this section: {\em the presence of bumps or a flattening in the positron spectrum may indicate the presence of a second, lighter WIMP}. Although no such features have been found at this point, future data releases may show them as PAMELA goes to higher energies and with the higher statistics and energies of AMS-02 \cite{Aguilar:2002ad}.\footnote{Note that the parameters chosen here, as in the examples through this paper, are consistent with the CMB limits described in \cite{Padmanabhan:2005es,Galli:2009zc,Slatyer:2009yq}.}

Such ``bumpy'' features have been considered before in the context of pulsars \cite{Profumo:2008ms,Malyshev:2009tw,Kawanaka:2009dk}, where bumps are associated with the duration between the initial release of positrons from the pulsar wind nebula and the present time. Although it is likely difficult to get some of the truly ``spikey'' signals considered by \cite{Profumo:2008ms,Malyshev:2009tw,Kawanaka:2009dk}, it is intriguing that broader features can be realized quite naturally for annihilating WIMP models, and are even expected in theories with multiple WIMPs (see additionally \cite{Frampton:2009yc} for a discussion of such features from multi-WIMP setups in decaying models) as well as pulsars.

\begin{figure}[h]
\centering
\begin{tabular}{cc}
\epsfig{figure=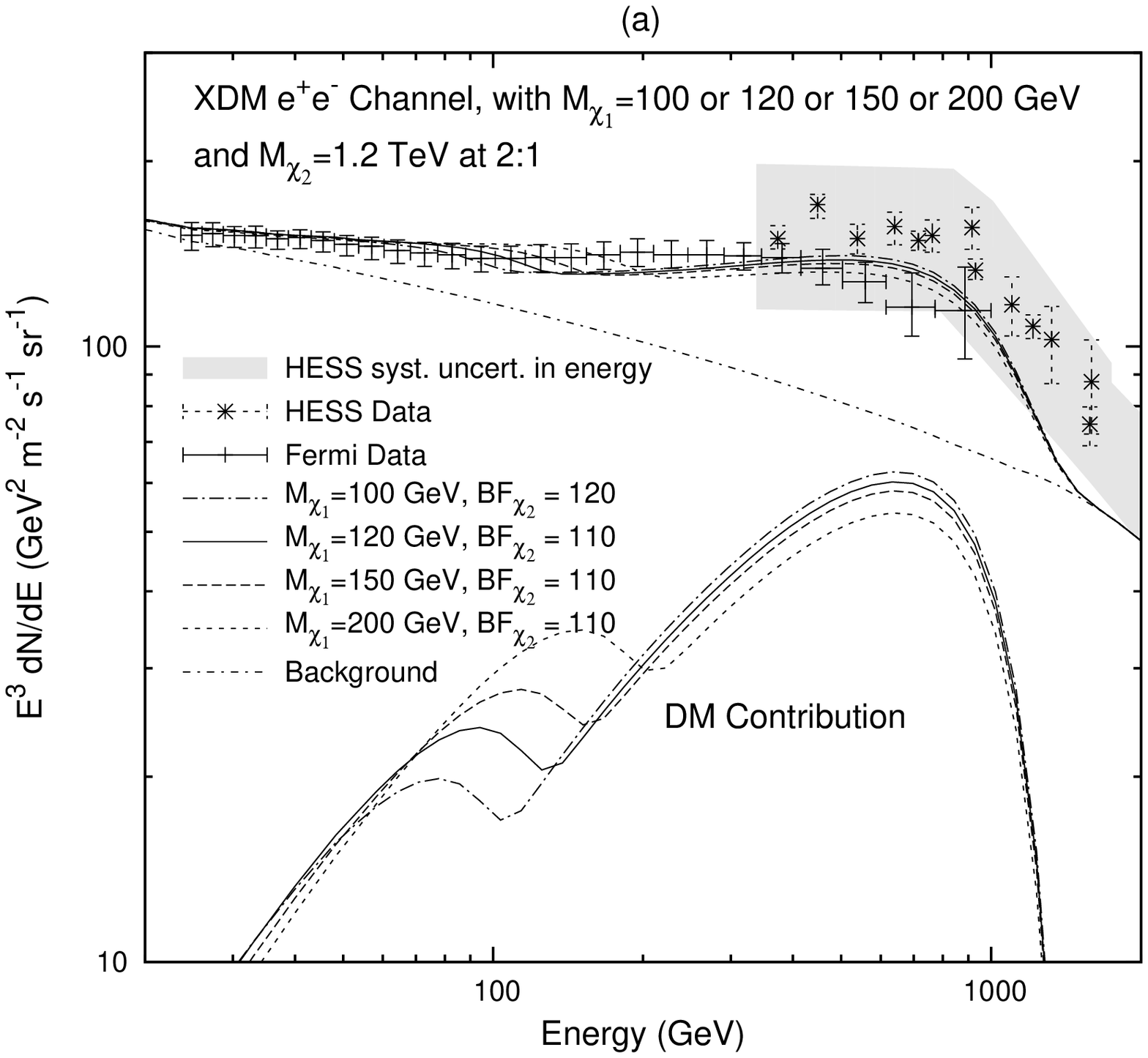,width=3.0in} &
\epsfig{figure=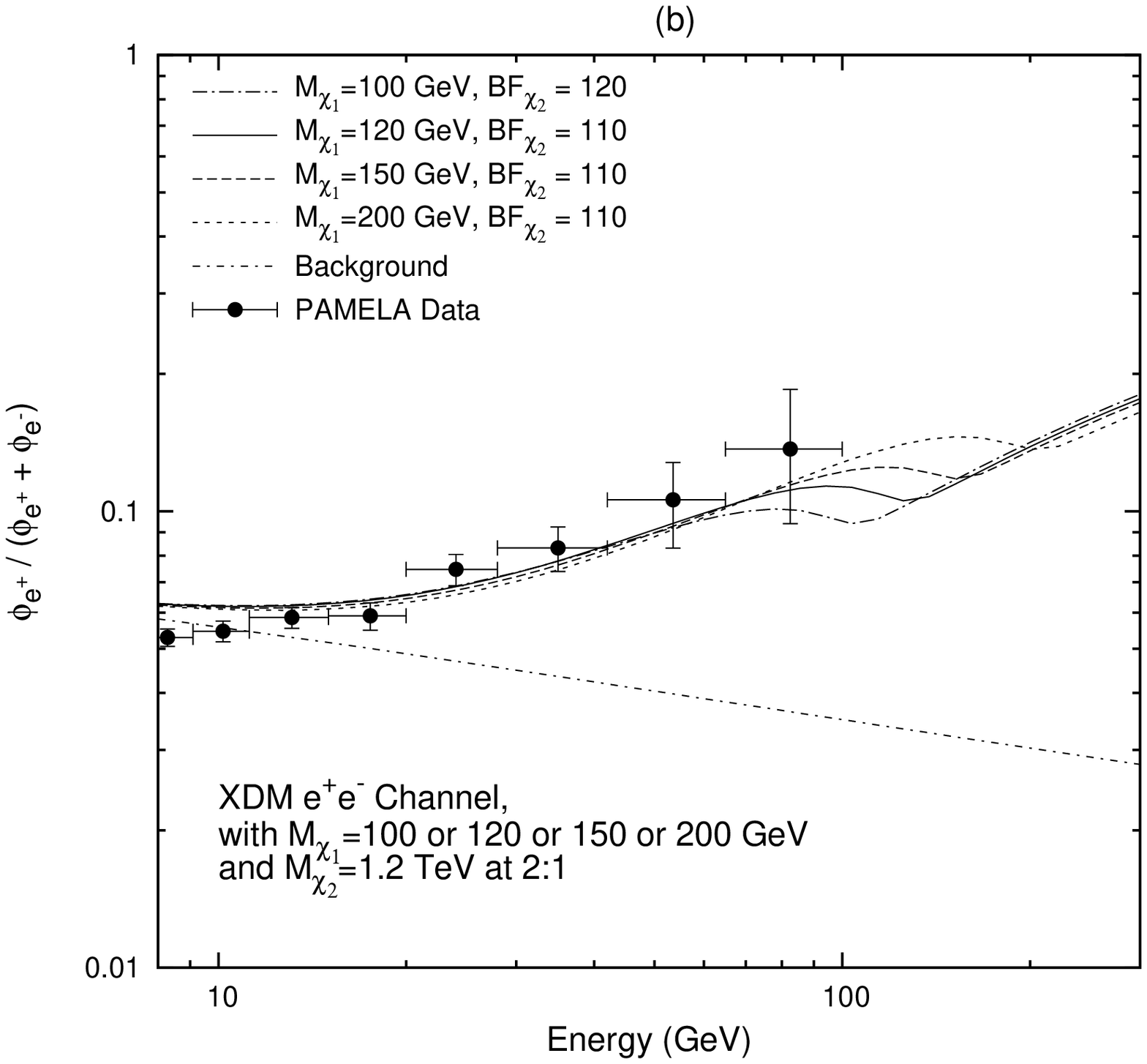,width=3.0in} 
\end{tabular}
\caption{Cosmic ray signals as in Fig. \protect \ref{fig:WIMPspectra}, with $M_{\chi_{1}}=100,120,150,200$ GeV and $M_{\chi_{2}}=1.2$ TeV through the XDM $e^{\pm}$ channel.}
\label{fig:WIMPspectramassvar}
\end{figure}

While in figure \ref{fig:WIMPspectra} we have varied the relative annihilation rates slightly, we can also vary the mass of the lighter state, which we do in figure \ref{fig:WIMPspectramassvar}. In cases where the lighter WIMP has a slightly larger cross section (plotted 2:1) the ``feature''  produced in the \epp spectrum would seem in tension with the Fermi data if it is too high $(\gsim 150\gev)$. Lighter WIMPs seem to have a good fit with the Fermi data, and it is intriguing to speculate on the change in slope in the experiment Fermi electron data. This slope change, while not statistically significant, could arise from the break associated with a $\sim 100$ GeV WIMP into light force carriers. Should this persist in the future, it may be pointing to the MiXDM scenario. 

\subsection{Annihilations with Different Force Carriers}
As noted above, it is possible to have different WIMPs, each with its own $\phi_i$. It is natural to expect in this case that the $\phi_i$ would receive different radiative corrections to their masses, which could naturally yield a situation where $M_i:\delta_i:m_i$ was constant.

As a means to study this, we will consider the situation where the lighter particle annihilates as $2 \chi_{1}$ $\rightarrow$ $2 \phi_{1}$ and then the mediator decays to leptons $\phi_{1}$ $\rightarrow$ $ l^{+}l^{-}$, while the heavier annihilates to $2 \chi_{2}$ $\rightarrow$ $2 \phi_{2}$ where  $\phi_{2}$ $\rightarrow$ $2 \phi_{2}'$ that $\phi_{2}'$ then decays to $l^{+}l^{-}$. For the analytical formulas for the injection spectra of the $e^{\pm}$ one can see\cite{Mardon:2009rc,DarkDiskpaper}.

In Fig.~\ref{fig:MXDM_1step_2step} we show the results where $l=e$ or $l=\mu$. This results in a very smooth change of the $e^{+} + e^{-}$ flux at $E=M_{\chi_{1}}$ relative to that of just $\chi_{2}$, for cases where relative annihilation rates of 1:1 are assumed. Both cases provide good fit to the Fermi data while their projected (B.F. calculated from fitting only to the Fermi data) positron fraction agrees with the PAMELA data.

\FIGURE[h]{
\centering
\begin{tabular}{cc}
\epsfig{figure=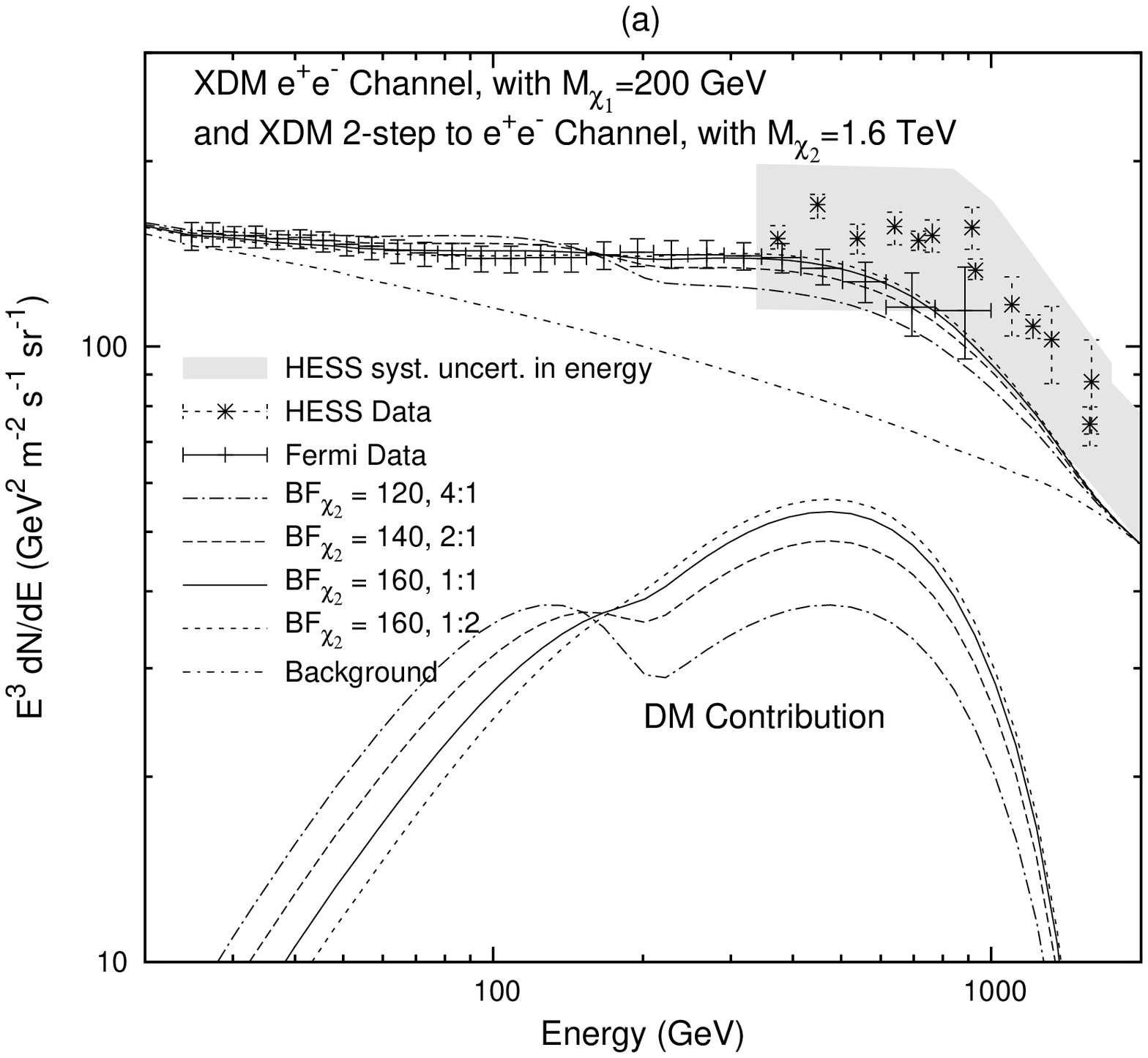,width=2.6in} &
\epsfig{figure=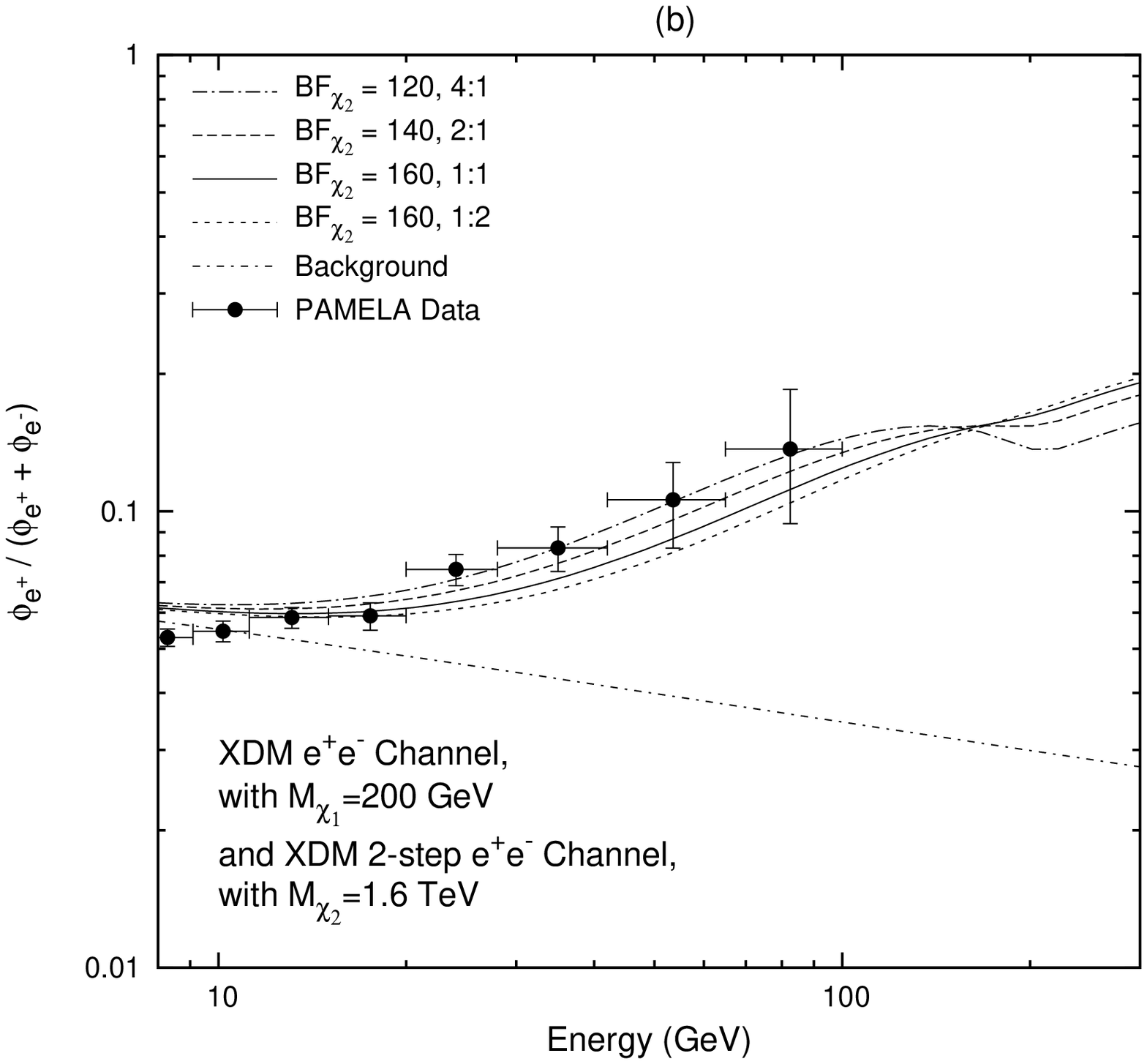,width=2.6in} \\
\epsfig{figure=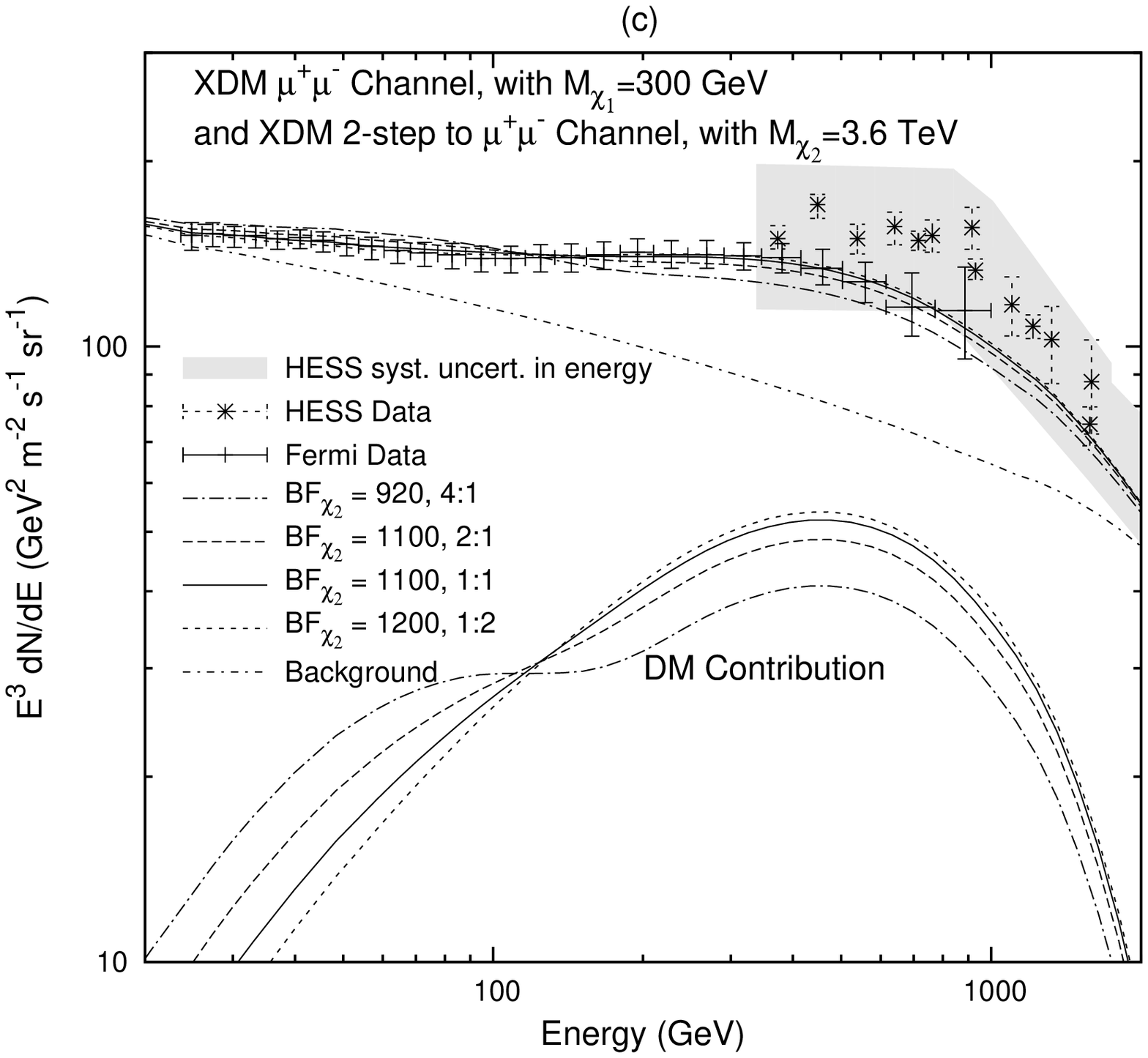,width=2.6in} &
\epsfig{figure=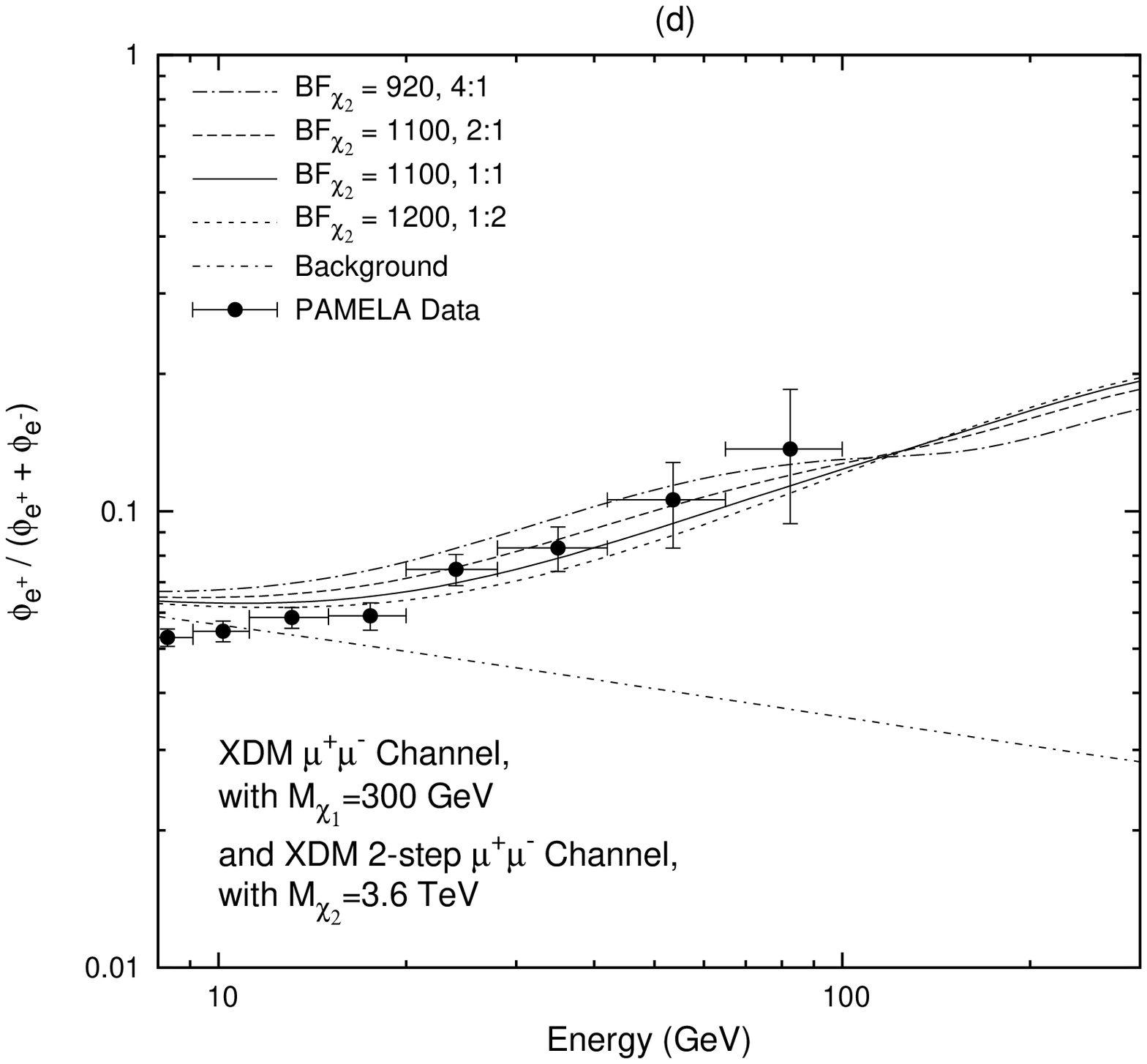,width=2.6in} \\
\epsfig{figure=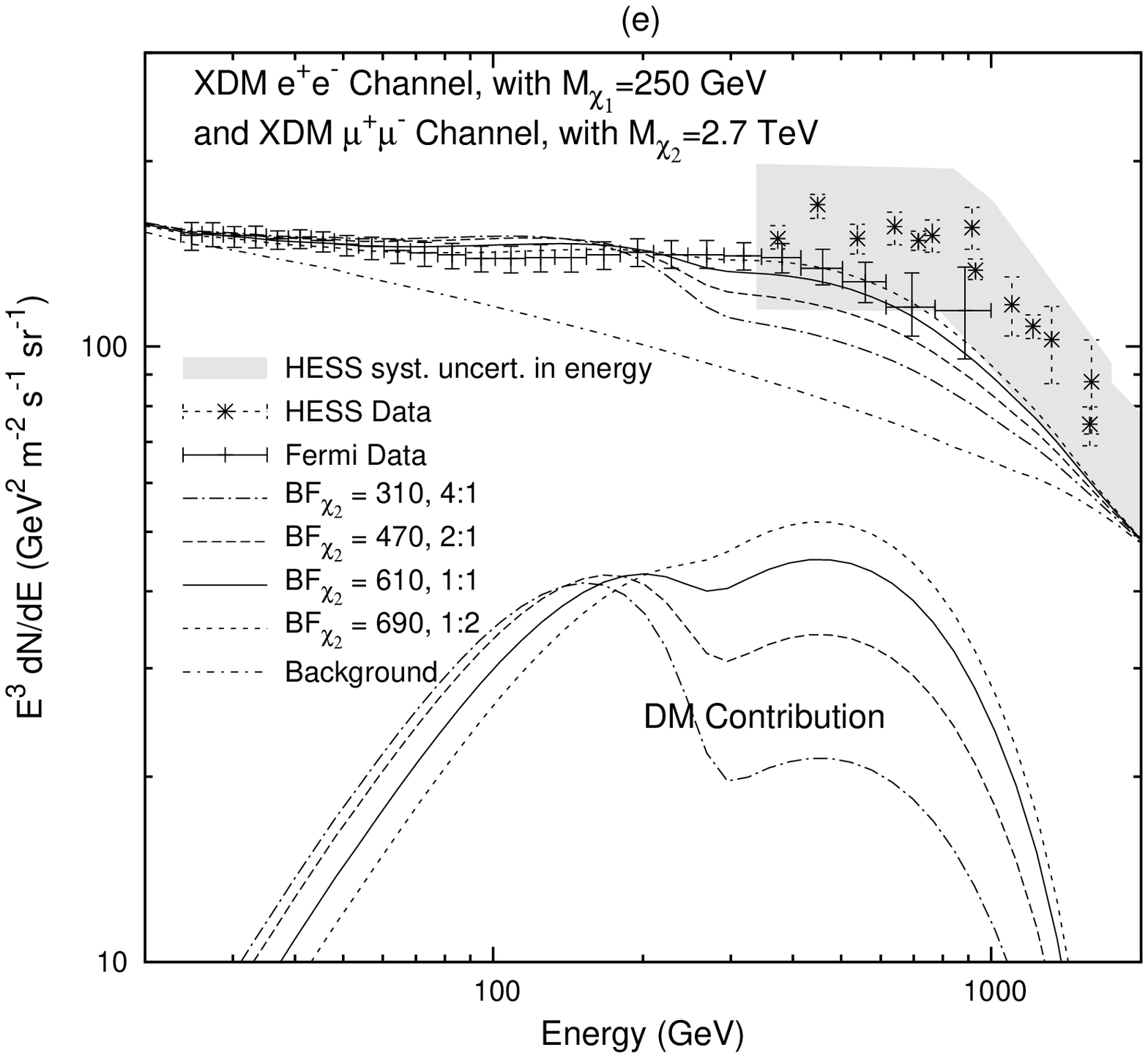,width=2.6in} &
\epsfig{figure=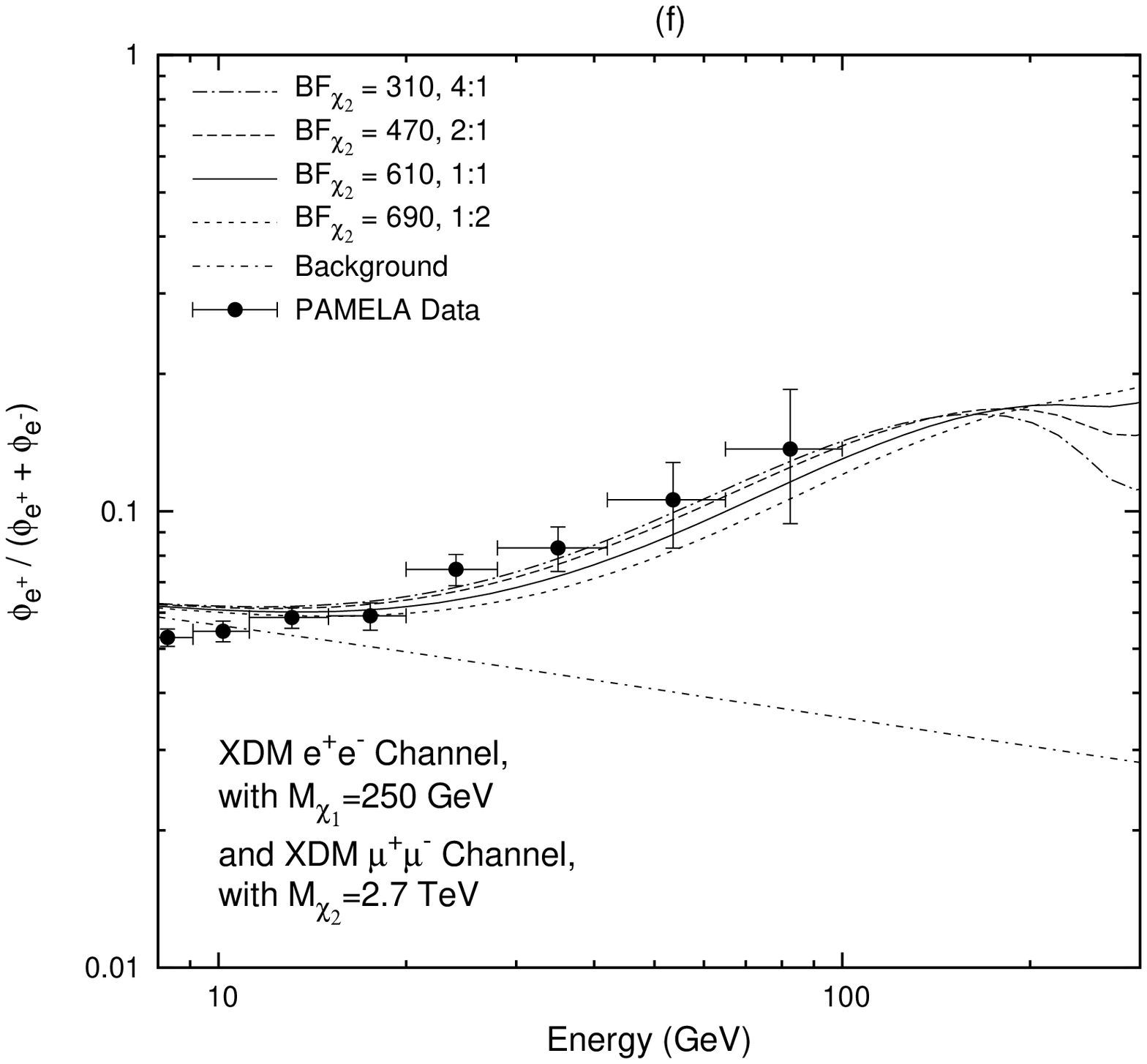,width=2.6in} 
\end{tabular}
\caption{Cosmic ray signals as in Fig. \protect \ref{fig:WIMPspectra}. Top: XDM $e^{\pm}$ with $M_{\chi_{1}}=200$ GeV and XDM 2-step to $e^{\pm}$ with $M_{\chi_{2}}=1.6$ TeV. Middle: XDM $\mu^{\pm}$ with $M_{\chi_{1}}=300$ GeV and XDM 2-step to $\mu^{\pm}$ with $m_{\chi_{2}}=3.6$ TeV. Bottom: XDM $e^{\pm}$ with $m_{\chi_{1}}=250$ GeV and XDM $\mu^{\pm}$ with $m_{\chi_{2}}=2.7$ TeV. }\label{fig:MXDM_1step_2step}
}
In Fig.~\ref{fig:MXDM_1step_2step}a,b we show the case where $M_{\chi_{1}}=200$GeV annihilating to 2$\phi_{1}$ with $m_{\phi_{1}}<2m_{\mu}$, thus $\phi_{1} \rightarrow e^{+}e^{-}$ (XDM $e^{+}e^{-}$ channel) and $M_{\chi_{2}}=1.6$TeV annihilating to 2$\phi_{2}$ where $\phi_{2} \rightarrow 2\phi_{2}'$ with $m_{\phi_{2}}\simeq 10m_{\phi_{2}'}$ and $\phi_{2}' \rightarrow e^{+}e^{-}$ as $m_{\phi_{2}'}<2m_{\mu}$(XDM 2-step $e^{+}e^{-}$ channel). In Fig.~\ref{fig:MXDM_1step_2step}c,d we study the case where $M_{\chi_{1}}=300$GeV and $M_{\chi_{2}}=3.6$TeV with $2m_{\mu} < m_{\phi_{1}}, m_{\phi_{2}'} < 2m_{\pi}$, thus the final product of the annihilations of the two species are $\mu^{+}\mu^{-}$ ($\phi_{1}, \phi_{2}' \rightarrow \mu^{+}\mu^{-}$), (XDM $\mu^{+}\mu^{-}$ channel for the lighter and XDM 2-step $\mu^{+}\mu^{-}$ channel for thr heavier species). Finally in Fig.~\ref{fig:MXDM_1step_2step}e,f we study the case where $M_{\chi_{1}}=250$GeV annihilating through XDM $e^{+}e^{-}$ channel and  $M_{\chi_{2}}=2.7$TeV through XDM $\mu^{+}\mu^{-}$ channel. Our choices for $M_{\chi_{2}}$ are suggested by the need to fit to the Fermi data and not violate the HESS upper limits on the $e^{+}e^{-}$ flux. As 2-body cascade decays of $\phi$'s and 3-body decays of muons produce significantly softer injection spectra of $e^{\pm}$ than XDM $e^{+}e^{-}$, higher masses of $M_{\chi_{2}}$ are needed to fit to the Fermi data.

In viewing the various examples of MiXDM annihilation signals, there are a few important issues that present themselves: first, while 4:1 rates typically clearly exhibit a feature in the electron and/or positron spectra, they are also generally in strong tension with the current Fermi data as well. Signals with 2:1 and smaller ratios are more difficult to distinguish in the electronic spectra. At the same time, the projected positron fraction fits the PAMELA data well for ratios of 1:1 and 2:1, alleviating some existing tension between the suggested annihilation rates from Fermi vs PAMELA data in single WIMP scenarios. Finally, should such a signal be present in the upcoming experimental results, it will still be unclear whether this is arising from dark matter, or from a conventional astrophysical source, such as pulsars. To answer the question whether a second WIMP is present in the data, we seek an additional means to constrain the signal.

\subsection{Inverse Compton Signals at Fermi}
As has been discussed elsewhere \cite{Cholis:2008wq,Zhang:2008tb,Borriello:2009fa,Cirelli:2009vg,Regis:2009md,Belikov:2009cx,Meade:2009iu}, models of WIMPs that explain PAMELA/Fermi also yield a substantial gamma ray signal in the galactic center region from inverse-Compton scattering. Remarkably, based upon the Fermi data release \cite{Dobler:2009xz} have argued that such a signal is present. Because MiXDM models have an additional soft \epp component, this can modify the spectrum of gamma rays.

\begin{figure}[t!]
\centering
\begin{tabular}{cc}
\epsfig{figure=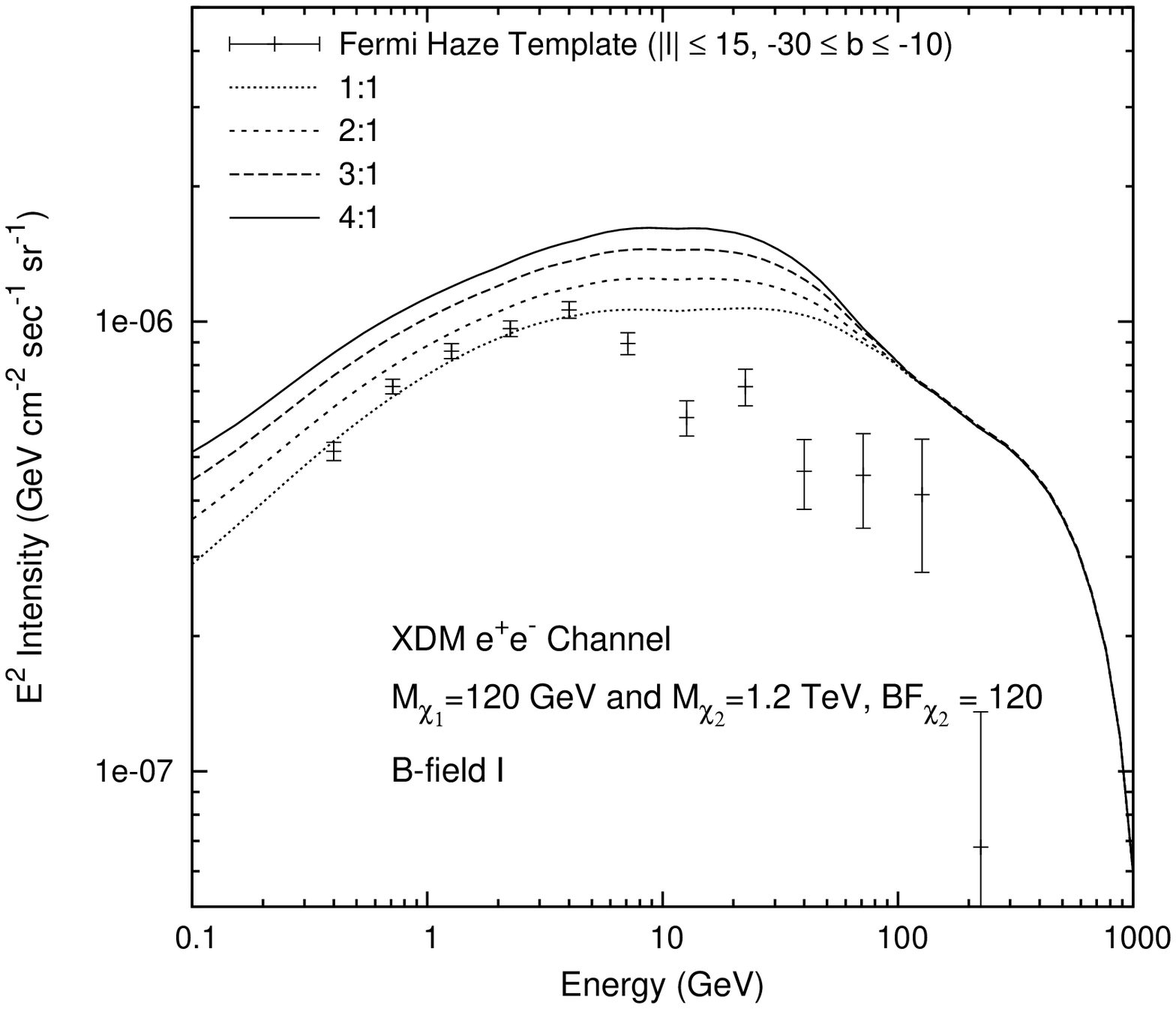,width=2.8in} &
\epsfig{figure=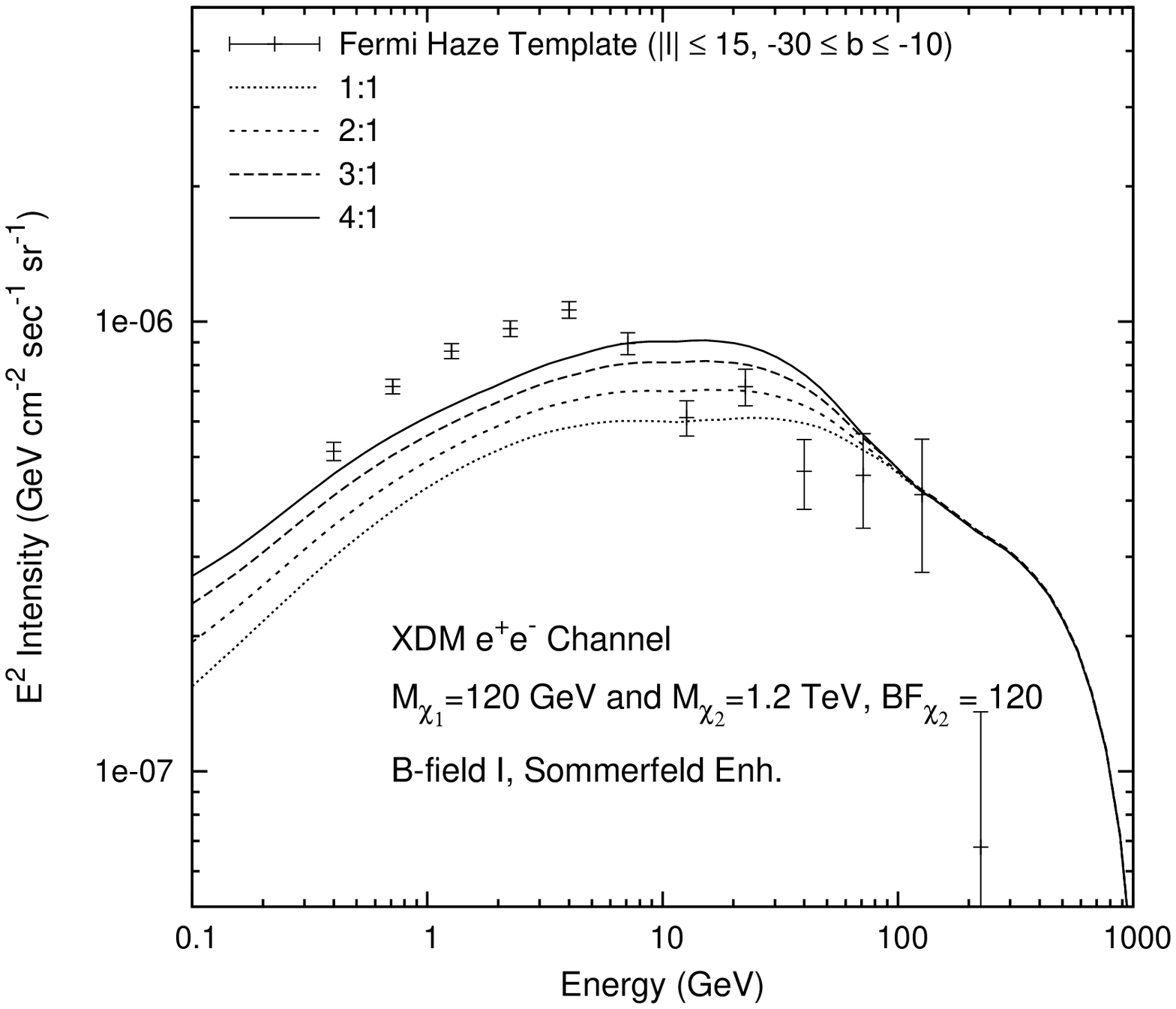,width=2.8in} \\
\end{tabular}
\caption{A comparison of the residual Fermi Haze data $|l|<10^\circ$ and $10^\circ < |b| < 30^\circ$ from \cite{Dobler:2009xz}, versus DM ICS contributions for uniform boost and an Einasto profile ({\it left}) and with a $v^{-1}$ scaling assuming $v_{disp} \propto r^{-1/4}$ ({\it right}).} 
\label{fig:icsplot}
\end{figure}

We show an example of the gamma ray spectrum for a MiXDM model in Fig \ref{fig:icsplot}. A somewhat smaller boost in the Haze region seems to give an additional contribution to the gamma rays which is at high energies comparable to what is extracted in \cite{Dobler:2009xz}. Such a variation could be easily accounted for by a $v^{-1}$ scaling, as shown.

\subsection{General fits and constraints}
In addition to the examples presented, we have also studied a variety of combinations of XDM $e^{\pm}$, XDM $\mu^{\pm}$, XDM $\pi^{\pm}$, XDM 2-step to $e^{\pm}$, XDM 2-step to $\mu^{\pm}$ and combinations of XDM to $e^{\pm}$ and $\mu^{\pm}$, with various annihilation ratios.  In general, for each of these channels we could find a combination of masses $M_{1}$ and $M_{2}$, that fit the Fermi $e^{+}+e^{-}$ flux data within 95$\%$ C.L. Thus - and not surprisingly - such combinations of two annihilating species fit the Fermi $e^{+}+e^{-}$ data to 95$\%$ C.L. in a fairly generic manner.

While we have demonstrated good agreement between the Fermi/PAMELA data sets from MiXDM models, we should consider other constraints as well. A variety of constraints from the diffuse gamma-ray background \cite{CyrRacine:2009yn,Profumo:2009uf} can arise, but the broad uncertainties in halo models allow the models discussed here to be unconstrained. In contrast, there are limits from the linear regime of the universe and the local halo that more directly connect to the local observations, or do so with fewer uncertainties about galaxy formation.

The most model-independent of these come from the CMB. 
Recently, \cite{Padmanabhan:2005es,Galli:2009zc,Slatyer:2009yq} have shown the CMB can constrain the maximum annihilation into $e^+e^-$ in the early universe.  At $z \sim 10^{3}$ those electrons will heat and ionize the photon-baryon plasma, changing the power spectrum and the polarization of the CMB. Thus constraints can be put on the annihilation rate at $z \sim 10^{3}$ based on the WMAP5 data.

Considering that the velocity dispersion of the DM at $z \sim 10^{3}$ is lower than what we expect locally, those constraints on the DM annihilation can be considered as the saturated values of the Sommerfeld enhancement (although note \cite{Slatyer:2009vg}).
\cite{Slatyer:2009yq} have calculated the maximum values of allowed DM annihilations for the XDM $e^{\pm}$, XDM $\mu^{\pm}$ and XDM $\pi^{\pm}$ channels. In all the cases presented in Fig.~\ref{fig:WIMPspectra} to~\ref{fig:MXDM_1step_2step} the suggested B.F. are below or similar to the maximum allowed Sommerfeld enhancement allowed by the WMAP5 data. Although there are some additional uncertainties in relating the local signal to that of the early universe (in particular, the local density and contributions from any local substructure), future data from Planck should constrain further the various annihilation channels \cite{Slatyer:2009yq}.

Dark matter annihilation into charged particles naturally produces high energy $\gamma$'s from final state radiation (FSR) which can place strong constraints in the galactic center (GC) and galactic ridge (GR) \cite{Bergstrom:2004cy,Birkedal:2005ep,Mack:2008wu}. These limits have been shown to be generally constraining for models that explain the Fermi electron spectrum \cite{Cirelli:2008pk,Bergstrom:2008ag,Meade:2009rb,Mardon:2009rc,Meade:2009iu,Fortin:2009rq}. However, this involves an extrapolation into the galactic center, which does involve uncertainties. As a simple example, we include an effect that has been heretofore ignored, namely, the variation in the velocity dispersion $v_{disp}$ as one approaches the galactic center. As noted earlier, many recent simulations with baryons have seen an increase in $v_{disp}$ roughly as $r^{-1/4}$  \cite{fabioprivate,RomanoDiaz:2008wz,RomanoDiaz:2009yq,Abadi:2009ve,Pedrosa:2009rw} as one moves to the inner galaxy. As the annihilation enhancement from the long-range force scales as $v^{-1}$ or $v^{-2}$ in resonance regions, this can suppress the signal in the GC and GR.

\begin{figure}[t!]
\centering
\begin{tabular}{cc}
\epsfig{figure=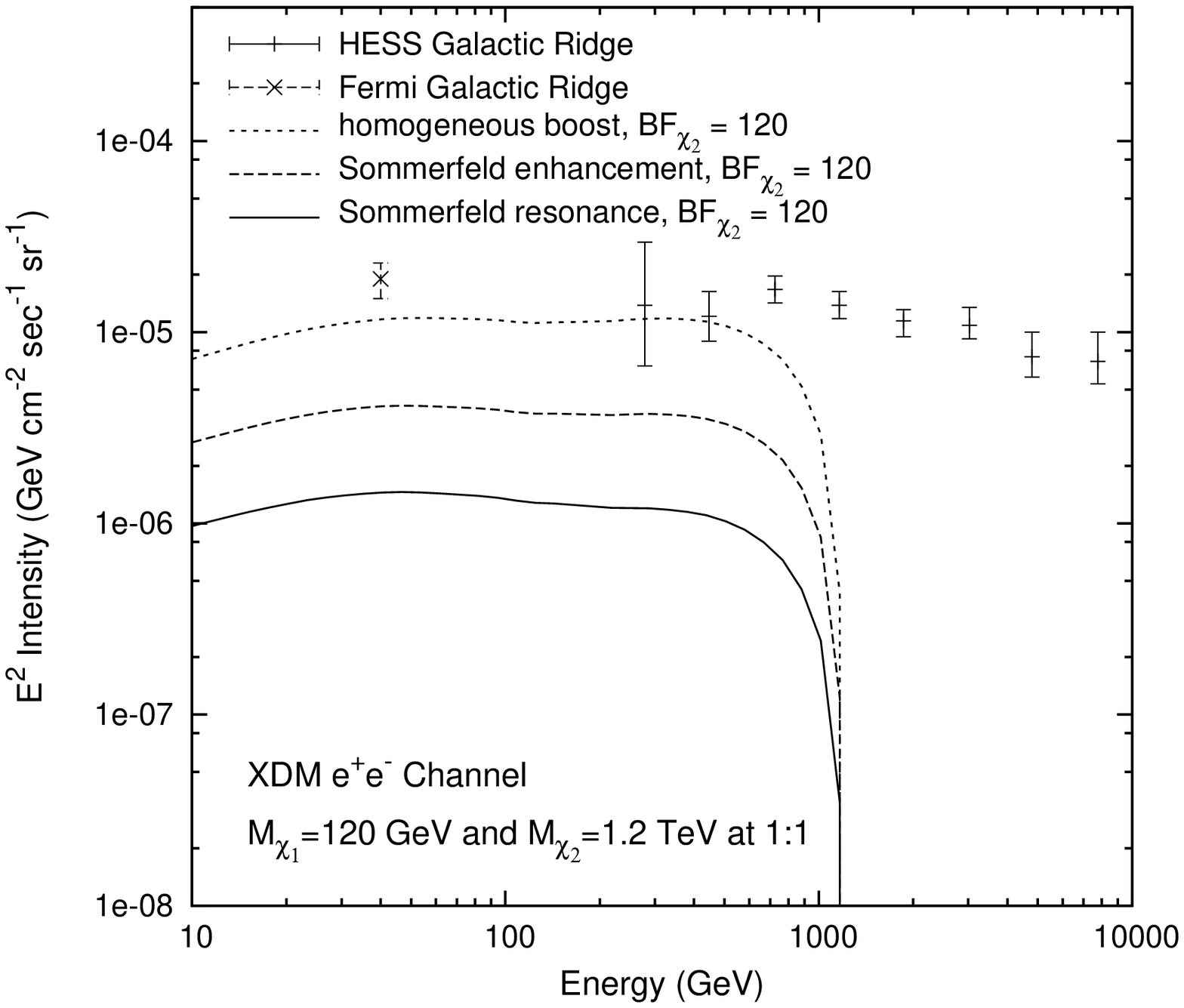,width=2.8in} 
\end{tabular}
\caption{A comparison of the signal from FSR+ICS versus HESS data in the Galactic Ridge, normalized to local PAMELA and Fermi signals assuming a cross section scaling as $v^0$, $v^{-1}$ and $v^{-2}$ with $v_{disp} \propto r^{-1/4}$, using an Einasto profile with $\alpha=0.17$. The lowest data point from Fermi is extracted from the maps of \cite{Dobler:2009xz}.} 
\label{fig:FSRGR}
\end{figure}

We show in figure Fig.~\ref{fig:FSRGR} a comparison between a MiXDM scenario normalized to fit the local cosmic ray data, under the assumption that the cross section is unchanged in the inner galaxy, or under the assumption that it has changed from the higher velocities. As is clear from the figure, these effects easily allow consistency with the HESS constraints. At the same time, the signal in the Haze region is only affected by a factor of 1.6 or 2.4 for $v^{-1}$ or $v^{-2}$, respectively. Of course, these curves assume extrapolation of a functional form for halo and velocity dispersion into regions that are not probed by simulations, and thus should be taken with a grain of salt\footnote{In this analysis, we are scaling the cross section as $\vev{\sigma v} \propto r^{-1/4}$ or $r^{-1/2}$. Once $\pi \alpha/v \sim 1$, this effect saturates. Even in the resonance case, this does not become relevant until the inner $\sim$ pc, and does not affect the plots shown.}.

\section{Counting the number of WIMP species with Synchrotron radiation}\label{sec:Haze}
      
One of the elements of the case for WIMP annihilation comes from the microwave excess in the galactic center, i.e. the ``microwave Haze'' \cite{Finkbeiner:2004us,Hooper:2007kb,Cholis:2008vb}. Such data have already been used to constrain the possible models that might explain it \cite{Hooper:2007kb}. Should Planck extend these data clearly to higher frequencies, it may serve to allow an additional handle on the details of WIMP annihilation.

To study this, we shall primarily use the simplified model shown in Fig.~\ref{fig:WIMPspectra} (XDM $e^{\pm}$ with $M_{\chi_{1}}=120$GeV and $M_{\chi_{2}}=1.2$TeV, with relative annihilations of 1:1.

\begin{figure}[t!]
\centering
\begin{tabular}{cc}
\epsfig{figure=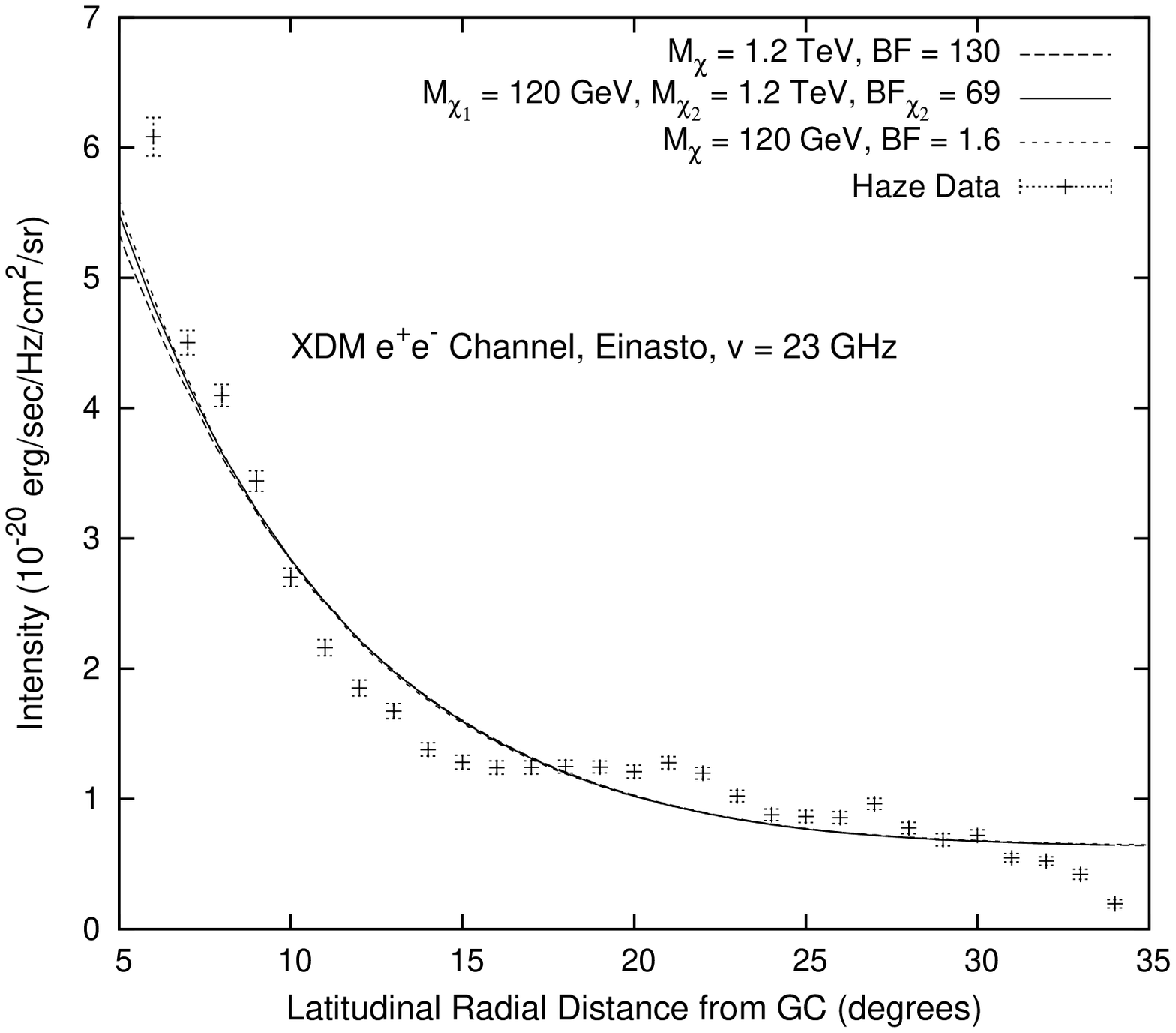,width=2.8in} \\
\end{tabular}
\caption{Synchrotron radiation at 22.5 GHz from $e^{\pm}$ of DM origin fitted to the microwave Haze of the same frequency. (\textit{Solid line}): flux from the two XDM $e^{\pm}$ species with masses 120 GeV and 1.2 TeV, and relative annihilation rates of 1:1.  (\textit{Dashed line}): fit of the 1.2 TeV only. (\textit{Dotted line}): equivalent fit for 120 GeV only.} 
\label{fig:Haze_23GHz}
\end{figure}

Before proceeding to the complete analysis, let us begin by studying the microwave Haze data we already have. As we show in Fig.~\ref{fig:Haze_23GHz} just using the 22.5 GHz WMAP band where the microwave Haze error bars are the smallest\footnote{not including the CMB bias}, it is not possible to distinguish between the cases where there is one mass of $M_{\chi}$=1.2 TeV from the MiXDM case of two 120 GeV / 1.2 TeV. This is not surprising, as $e^{\pm}$ that emit synchrotron radiation at 22.5 GHz have energies between 5 $\lsim$ to 50 GeV, so $e^{\pm}$ originating from both species contribute essentially equally.

\subsection{Modeling Synchrotron Emission with Power Laws}

\begin{figure}[t!]
\centering
\begin{tabular}{cc}
\epsfig{figure=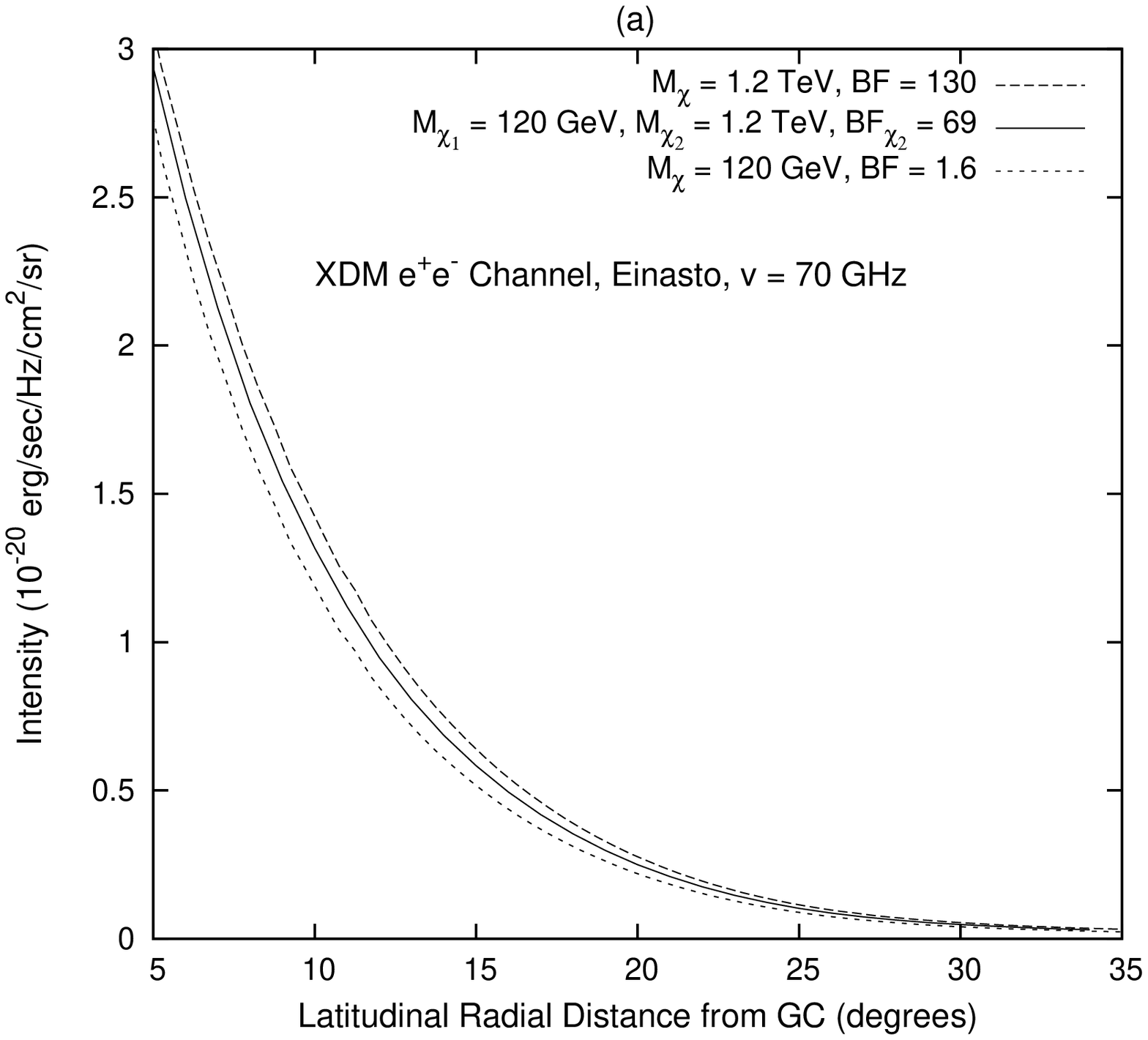,width=2.8in} &
\epsfig{figure=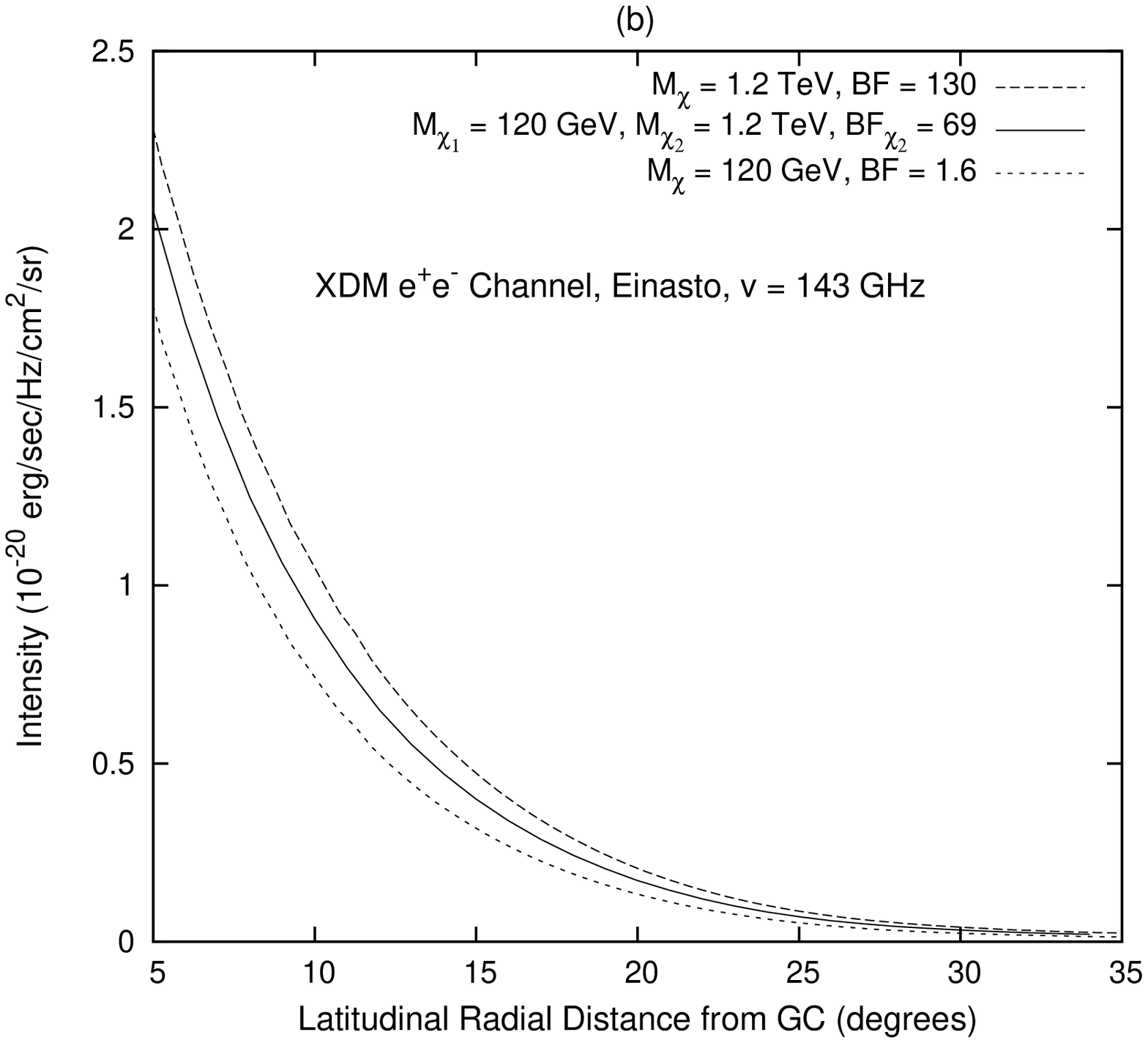,width=2.8in} \\
\vspace{3pt}
\epsfig{figure=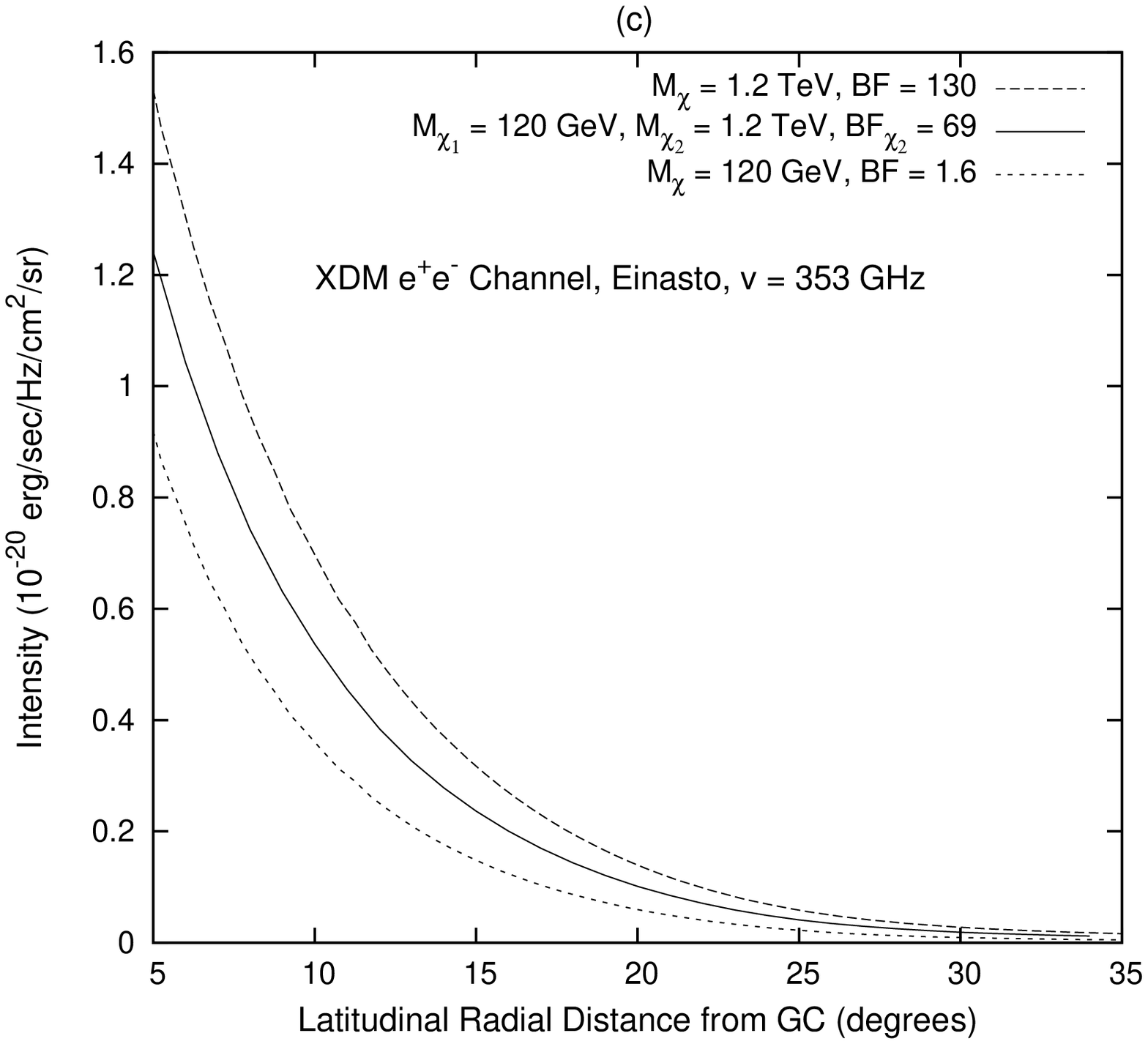,width=2.8in} &
\epsfig{figure=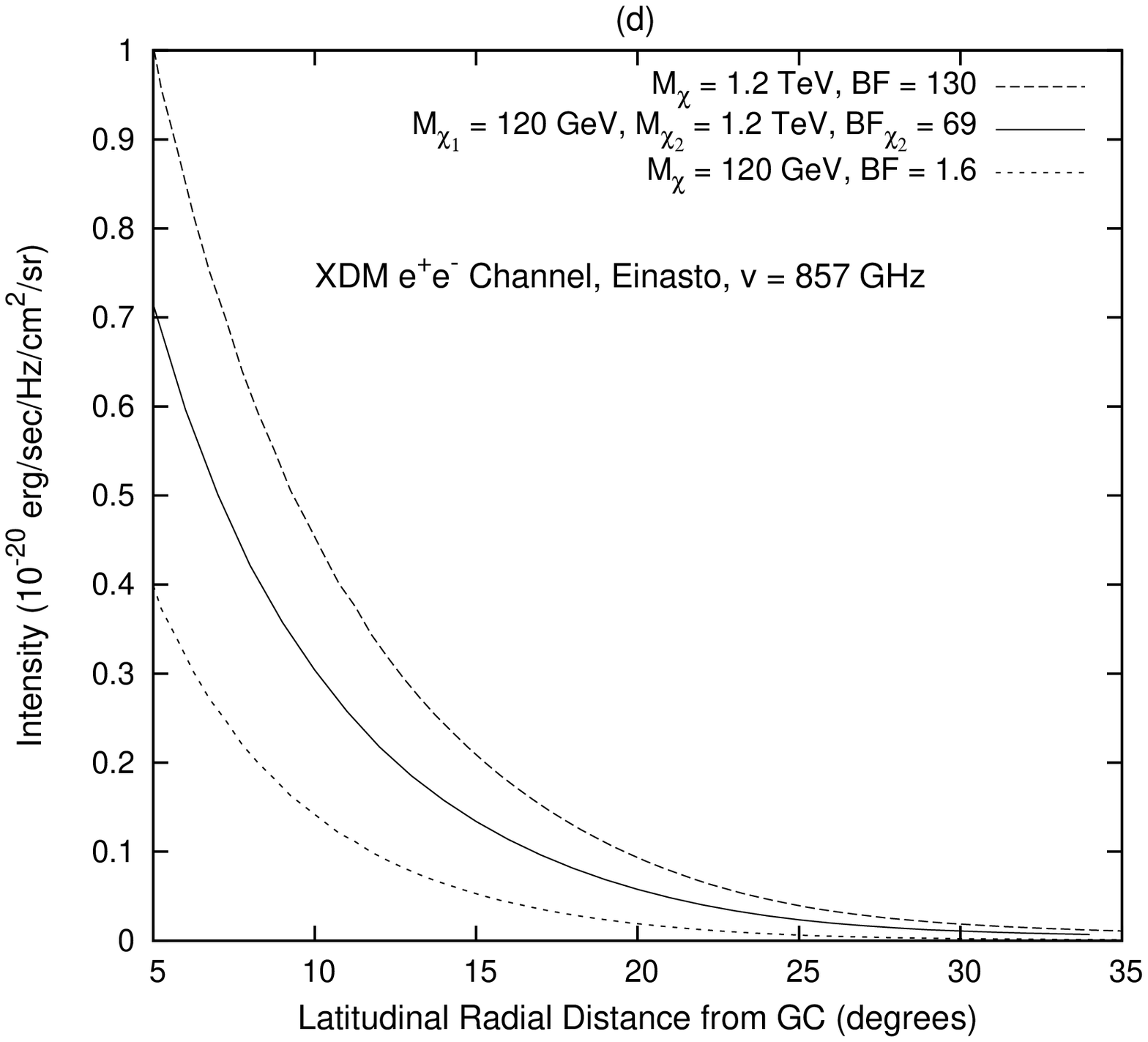,width=2.8in} \\
\end{tabular}
\caption{Synchrotron radiation from the $e^{\pm}$ of DM origin as in Fig. \protect \ref{fig:Haze_23GHz}, for 70 GHz (top left), 143 GHz (top right), 353 GHz (bottom left) and 857 GHz (bottom right). Boost factors are from fitting to the 22.5 GHz microwave Haze.}
\label{fig:HAZE_high_freq}
\end{figure}

As we discuss in the appendix, in considering the emissivity $j(\textrm{v})$ in the range of  $\sim 100$ GHz, one should expect the synchrotron emissivity $J_{1}(\textrm{v})$ of the full distribution of $e^{\pm}$ that originate from $\chi_{1}$ (with $M_{\chi_{1}}=120$ GeV) to be described by a steeper power law than that of $\chi_{2}$ (with $M_{\chi_{2}}=1.2$ TeV),
\begin{equation} 
J_{1}(\textrm{v}) \sim \textrm{v}^{-\alpha_{1}} \; , \; J_{2}(\textrm{v}) \sim \textrm{v}^{-\alpha_{2}} \; \textrm{with} \; \alpha_{1} > \alpha_{2},
\label{eq:synch_emiss_power_laws}
\end{equation}
simply because there is more energy in high energy electrons from $\chi_{2}$ to contribute to the higher frequencies.
That is also shown in Fig.~\ref{fig:HAZE_high_freq} where we present the synchrotron radiation from the $e^{\pm}$ of DM origin vs latitude for v: 70, 143, 353 and 857 GHz (frequencies to be measured by Planck). The calculated values of the synchrotron radiation are averaged over longitudes of $-10^{\circ}$ to $+10^{\circ}$ as the WMAP microwave Haze is calculated from that region of the sky \cite{Dobler:2007wv}. The annihilation cross sections for those particles are fixed by fitting to the 22.5 GHz WMAP microwave Haze shown in Fig.~\ref{fig:Haze_23GHz}. As explained, the synchrotron emissivity of $e^{\pm}$ from a 120 GeV DM particle drops faster than that of a 1.2 TeV DM particle. 

Note that in Fig.~\ref{fig:Haze_23GHz} we fit the 22.5 GHz microwave Haze band data with a combination of the synchrotron radiation from the $e^{\pm}$ of DM origin and an extra constant offset, as the overall zero of the microwave Haze is somewhat uncertain. In Fig.~\ref{fig:HAZE_high_freq} instead we present only the synchrotron radiation from the $e^{\pm}$ of DM origin, but it should be expected that  with actual data some constant offset will be required.

As a result of eq.~\ref{eq:synch_emiss_power_laws} the combination distribution in the 2-species case will be described by a power law $J_{comb}\sim \textrm{v}^{-\alpha_{comb}}$ of intermediate steepness.
\begin{equation} 
J_{comb}(\textrm{v}) \sim \textrm{v}^{-\alpha_{comb}} \; \textrm{with} \; \alpha_{1} > \alpha_{comb} > \alpha_{2} \; .
\label{eq:combined_synch_emiss_power_law}
\end{equation}

The key point in~(\ref{eq:combined_synch_emiss_power_law}), is that $\alpha_{1} \neq \alpha_{2}$. While this may appear obvious, there are actually subtleties involved that we shall explain. The obvious element is the direct relationship between synchrotron spectrum and the hardness of the electron spectrum. From synchrotron theory we know that for a power law distribution of electron energies of the form $N(E)=\kappa E^{-p}$, the synchrotron emissivity of the distribution goes as\cite{Longair}:
\begin{equation} 
J(\textrm{v}) \propto \kappa \textrm{v}^{-(p-1)/2} \; .
\label{eq:synch_emiss_power_law_from_elec_spect_power_law}
\end{equation}

Considering that for both masses we used the same annihilation channel, one would think that $J_{1}(\textrm{v})$ and $J_{2}(\textrm{v})$ should be described by the same power law, which is true - up to a point. The injection spectra are given by the same formula\cite{Cholis:2008vb} which gives an injection power law of 0, and the dominant energy losses (synchrotron and ICS) scale as $\frac{dE}{dt}\propto E^{-2}$. However, eq. (\ref{eq:synch_emiss_power_law_from_elec_spect_power_law})  neglects the energy cut-off $E_{cf}$ in the distribution of the electrons. 
As we explained earlier in this section, it is exactly the presence of the cut-off that for frequencies $\textrm{v} \not\ll \textrm{v}_{c_{cf}}$~(\ref{eq:synch_emiss_power_law_from_elec_spect_power_law}) where:
\begin{equation} 
\textrm{v}_{c_{cf}} = \frac{3}{2}\frac{eBE_{cf}^{2}}{2\pi m_{e}^{3}}\sin\alpha \; ,
\label{eq:char_freq_cut_off}
\end{equation}
shouldn't be applied blindly. 
Thus $\alpha_{1} \neq \alpha_{2}$. 

A second element contributing to $\alpha_{1} > \alpha_{2}$ is that the Thompson approximation of the cross-section of ICS from optical starlight breaks down at electron energies $E \sim 100$ GeV, with the actual cross section between the electron and the photon being described by the Klein-Nishina cross section  \cite{Longair}: 
\begin{equation}
\sigma_{K-N} = \frac{3}{8}\sigma_{T}x^{-1}\left\{\left[1-\frac{2(x+1)}{x^{2}}\right]\ln(2x+1)+\frac{1}{2}+\frac{4}{x}-\frac{1}{2(2x+1)^{2}}\right\},
\label{eq:Klein_Nishina_cross_section}
\end{equation}
where $x=\frac{\hbar\omega}{m_{e}c^{2}}$. At the limit where $x \gg 1$ equation~\ref{eq:Klein_Nishina_cross_section} simplifies to:
\begin{equation}
\sigma_{K-N} = \frac{3}{8}\sigma_{T}x^{-1}\left(\ln(2x) + \frac{1}{2}\right).
\label{eq:Klein_Nishina_cross_section_approx}
\end{equation}
This results in an energy loss rate $\frac{dE}{dt}$ due to ICS from optical starlight that scales as $E^{\beta}$ with $\beta < 2$ and gradually decreasing with increasing electron energy $E$.
As a consequence, ICS is a less efficient energy loss for the higher energy electrons, resulting in more energy (fractionally) being lost to synchrotron radiation, and a harder synchrotron spectrum.

\begin{figure}[t!]
\centering
\begin{tabular}{cc}
\epsfig{figure=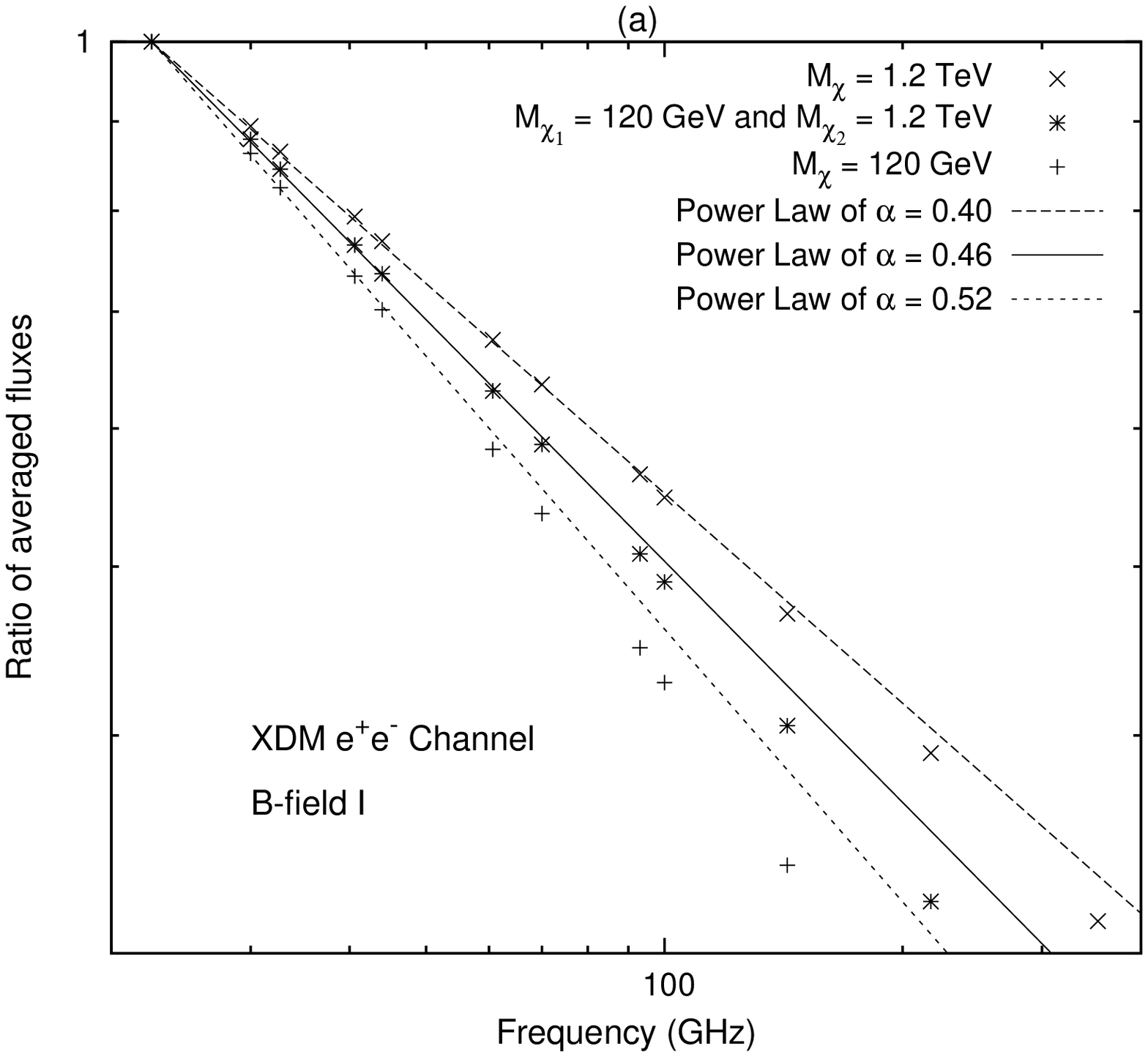,width=2.8in} &
\epsfig{figure=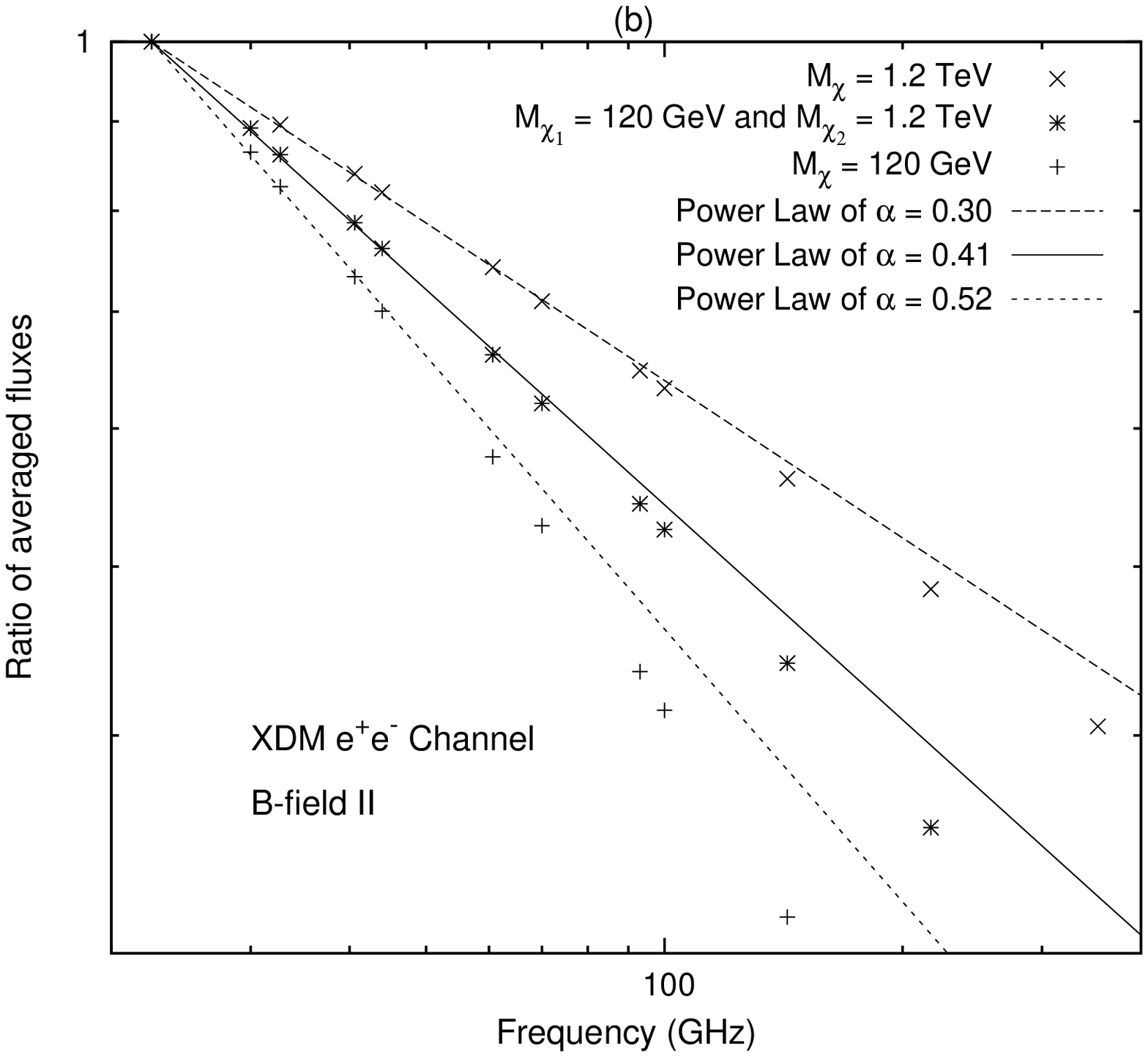,width=2.8in} \\
\end{tabular}
\vspace{3pt}
\begin{tabular}{c}
\epsfig{figure=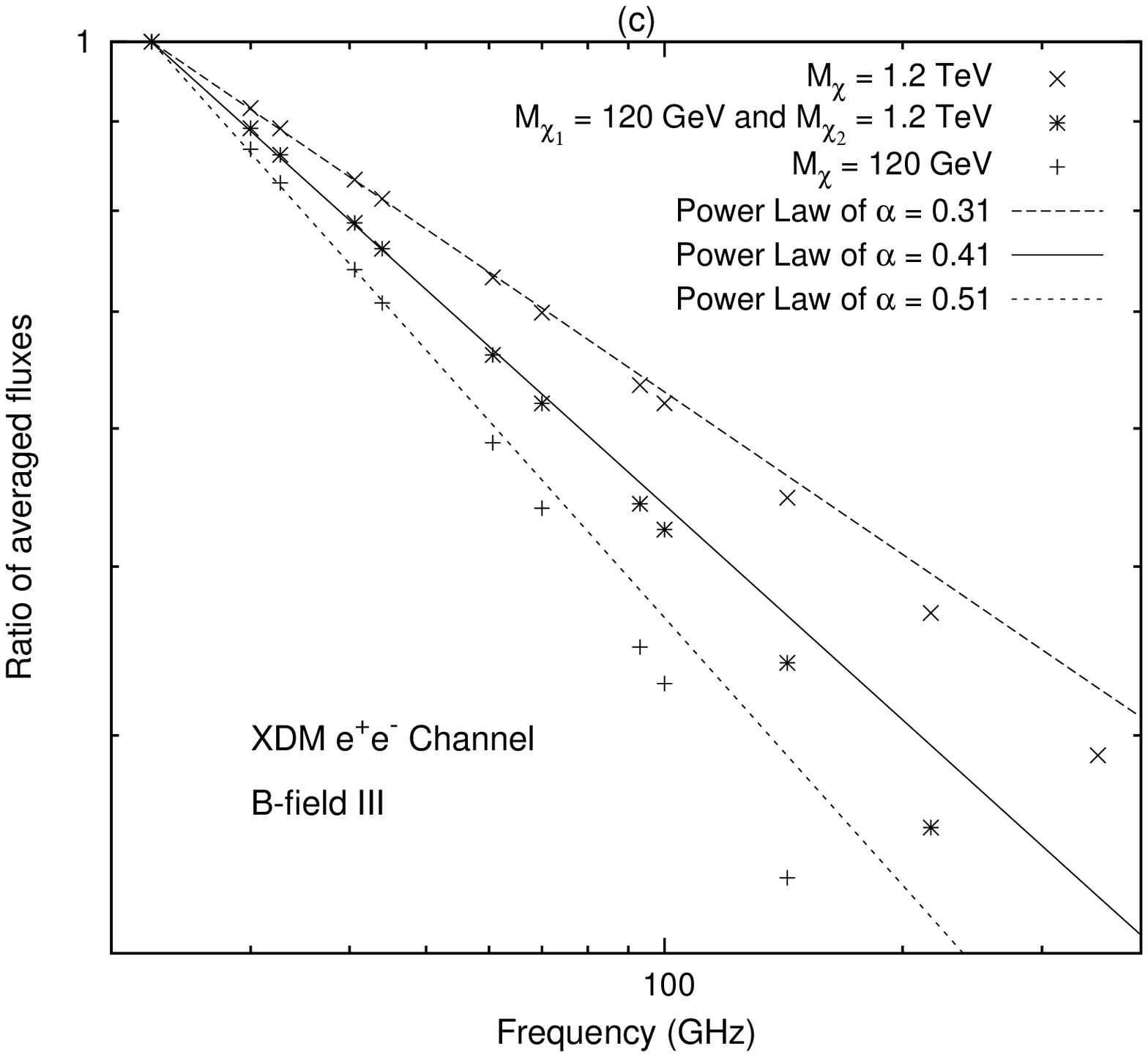,width=2.8in} \\
\end{tabular}
\caption{Plot (a): Ratio of synchrotron radiation at frequency v over radiation at 22.5 GHz from $e^{\pm}$ of DM origin. Assumed magnetic field model: ``B-field I'' (\textit{Solid line}): ratio for the two XDM $e^{\pm}$ species with masses 120 GeV and 1.2 TeV, at ($\chi_{1}$:$\chi_{2}$) of 1:1.  (\textit{Dashed line}): ratio for the 1.2 TeV only. (\textit{Dotted line}): equivalent ratio for 120 GeV only. Plot (b): Same as for (a) but with magnetic field model ``B-field II''. Plot (c): Same as (a) but with magnetic field model ``B-field III''.}
\label{fig:ratio_of_synch_rad}
\end{figure}

In Fig.~\ref{fig:ratio_of_synch_rad} we show the ratio of synchrotron radiation flux at frequency v averaged between $6^{\circ}$ to $15^{\circ}$ in latitude and $-10^{\circ}$ to $+10^{\circ}$ in longitude (microwave Haze region), to the same averaged flux at 22.5 GHz. 
\begin{equation} 
\frac{\int J(\textrm{v}) d\Omega dr}{\int J(\textrm{v=22.5 GHz}) d\Omega dr} \; ,
\label{eq:ratio_of_synch_rad}
\end{equation}
where $dr$ refers to the integration over the line of light and $d\Omega$ to the integration over the sky region. That ratio is calculated for the cases of XDM $e^{\pm}$ channel with $M_{\chi} = 120$ GeV, $M_{\chi} = 1.2$ TeV separately as well as the MiXDM case. We use the 5 WMAP frequencies: 22.5, 32.7, 40.6, 60.7 and 93.1 GHz \cite{Bennett:2003bz} and the 9 Planck frequencies 30, 44, 70 GHz (measured from the Low Frequency Instrument)\cite{Mandolesi:1999ba,Mennella:2003qp} and 100, 143, 217, 353, 545 and 857 GHz (measured from the High Frequency Instrument)\cite{Lamarre:2003zh,:2006uk}. We also show the indices of the power laws that fit our calculations from the lowest frequency (22.5 GHz) up to $\sim$ 100 GHz, where deviations from the one power law description arise.  

Based on (\ref{eq:combined_synch_emiss_power_law}) if we are able to measure the power law index $\alpha$ of $J(\textrm{v})$, and if we know the mass $M_{\chi}$(or $M_{\chi_{2}}$ for the 2-mass case) and the annihilation channel via which $e^{\pm}$ are produced then we can discriminate between the 2-species $\chi_{1}$, $\chi_{2}$ case and the case where there is only $\chi_{2}$ in nature. As shown in Fig.~\ref{fig:ratio_of_synch_rad}a\footnote{See section~\ref{sec:B-filed} for definitions of magnetic fields}, the power law for frequencies up to $\sim 100$ GeV, is for the case of a single species $M_{\chi} = 1.2$ TeV equal to $\alpha = 0.40$ while for the 2-species case with  $M_{\chi_{1}} = 120$ GeV and $M_{\chi_{2}} = 1.2$ TeV it is $\alpha = 0.46$, because of the contribution from $\chi_{1}$ that falls with a steeper power law ($\alpha = 0.52$). Also in Fig.~\ref{fig:ratio_of_synch_rad} notice that at v greater than $\sim$100 GHz, the power-law approximation of the $e^{\pm}$ distribution stops being valid.

In general, for the MiXDM scenario in which masses are separated by an order of magnitude or greater, the difference in the power law index compared to the single WIMP cases is $\simeq 0.1$ or greater. There are many uncertainties and challenges in extracting the Haze from future Planck data, so it remains to be seen whether this will be adequate. In this paper we do not intend to discuss the uncertainties arising from understanding the various astrophysical contributions (free-free emission, synchrotron radiation from primary electrons and thermal and spinning dust)\cite{Finkbeiner:2003im,Dobler:2007wv}. Currently only the 22.5 and 32.7 GHz from WMAP have small enough error bars to be used in order to calculate the power law $\alpha$. In \cite{Hooper:2007kb}, the authors find that the ratio of the averaged Haze flux at 22.5 GHz to that at 32.7 GHz is $J(22.5 GHz)/J(32.7 GHz)\,\approx \, 1.18 \pm 0.10$ that gives a range, $0.2 < \alpha < 0.6$.   
Since \cite{Hooper:2007kb} used only the statistical errors of the microwave Haze, while there are additional large uncertainties arising predominantly from the microwave Haze's correlation with the CMB. Thus the actual range of values of $\alpha$ from the microwave Haze is even greater than this. At the moment, one- and two- species scenarios are indistinguishable within errors.

\begin{figure}[t!]
\centering
\begin{tabular}{cc}
\epsfig{figure=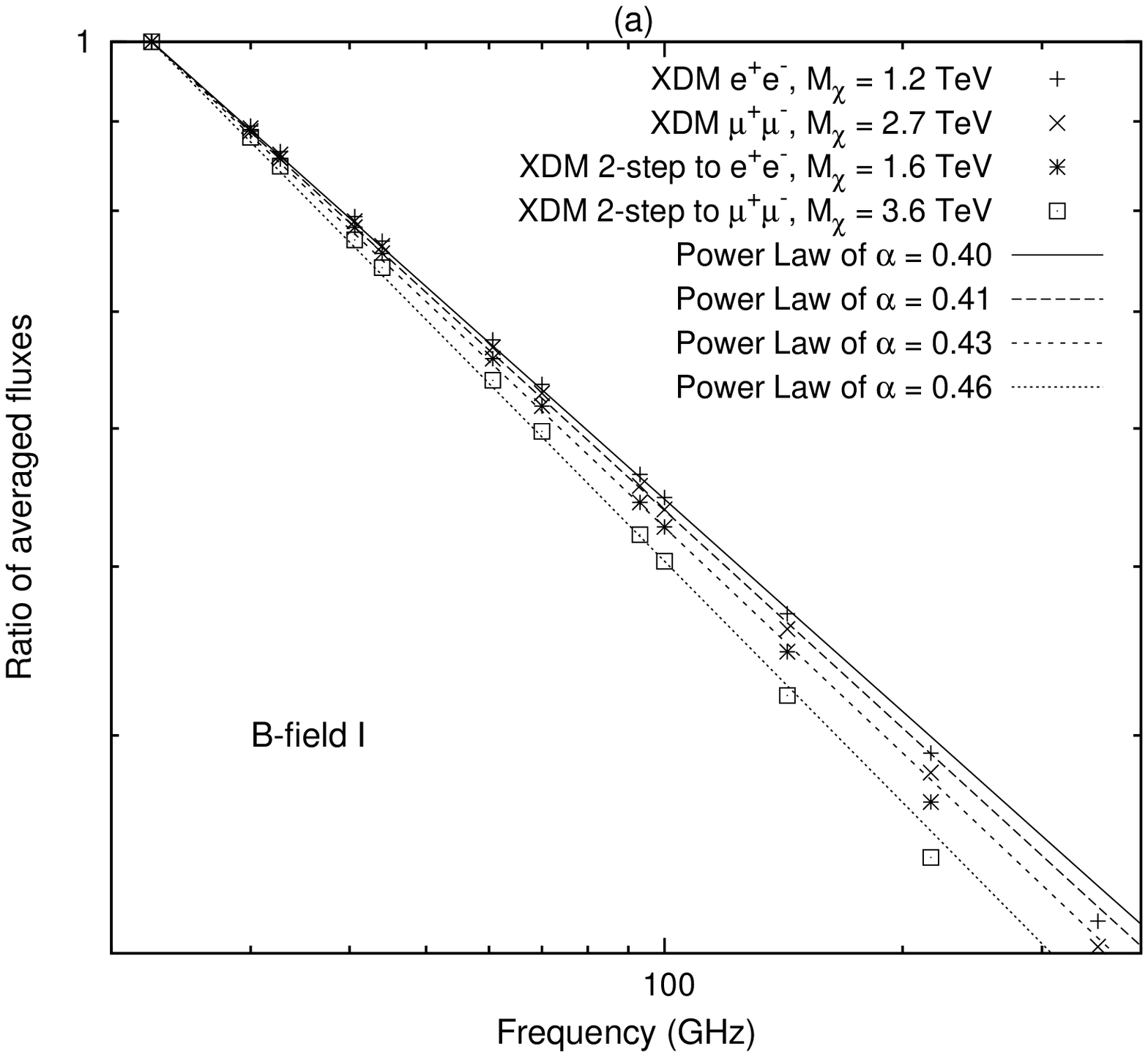,width=2.8in} &
\epsfig{figure=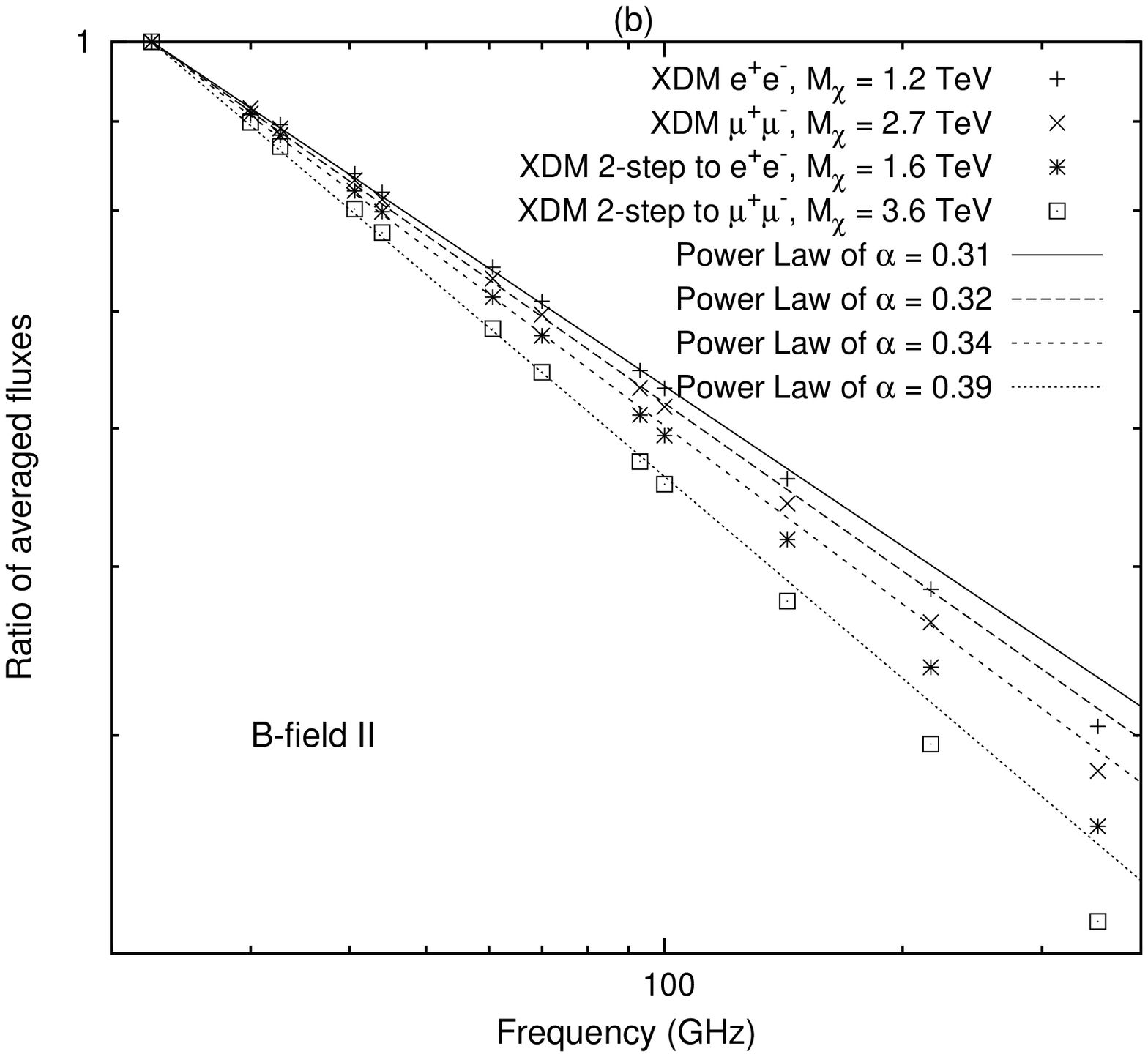,width=2.8in}\\
\end{tabular}
\vspace{3pt}
\begin{tabular}{c}
\epsfig{figure=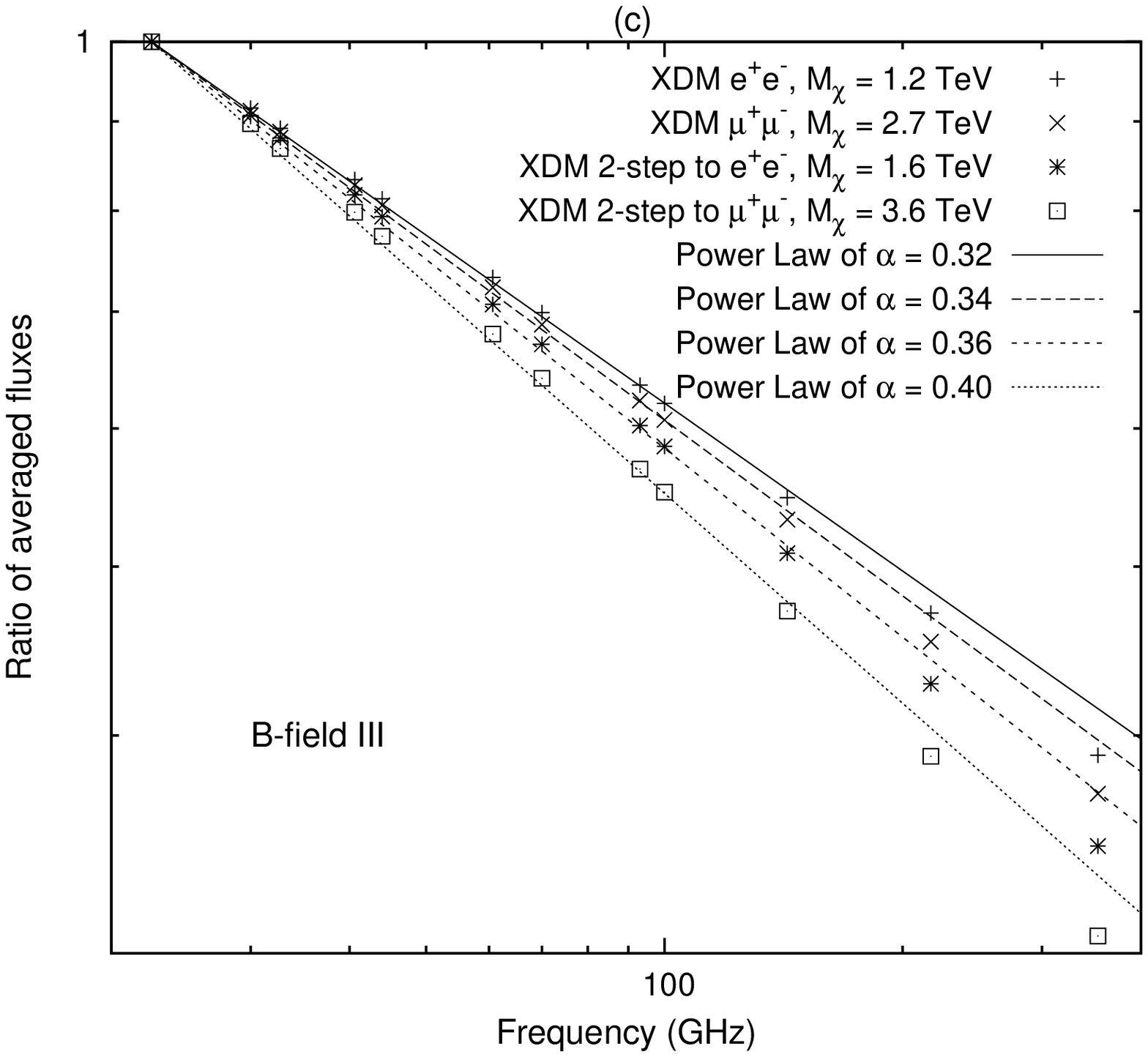,width=2.8in} \\
\end{tabular}
\caption{Plot (a): Ratio of synchrotron radiation as in \protect \ref{fig:ratio_of_synch_rad}a (used magnetic field model: ``B-field I''). (\textit{Solid line}): ratio for the XDM $e^{\pm}$ species with $M_{\chi} = 1.2$ TeV. (\textit{Long dashed line}): ratio for the XDM $\mu^{\pm}$ species with $M_{\chi} = 2.7$ TeV. (\textit{Short dashed line}): ratio for the XDM 2-step cascade to $e^{\pm}$ species with $M_{\chi} = 1.6$ TeV. (\textit{Dotted line}): ratio for the XDM 2-step cascade to $\mu^{\pm}$ species with $M_{\chi} = 3.6$ TeV. Plot (b): Same as in (a) with magnetic field model ``B-field II''. Plot (c): Same as in (a) with magnetic field model ``B-field III''.}
\label{fig:ratio_of_synch_rad_4channels}
\end{figure}

Another difficulty in determining whether we are in a MiXDM scenario is the uncertainty in the annihilation channel, itself. Constraints from fitting the Fermi/HESS electron data require $1 \tev \lsim M_{\chi_{2}} \lsim 4 \tev$, depending on the annihilation mode, with some significantly softer than others. Such changes will clearly impact the spectrum of the synchrotron emission $J(\textrm{v})$. To explore this, we show in Fig.~\ref{fig:ratio_of_synch_rad_4channels} the ratio of synchrotron radiation flux at frequency v (normalized to 22.5 GHz) averaged over the microwave Haze region (as in Fig.~\ref{fig:ratio_of_synch_rad}) for four different annihilation channels for single species of mass $M_{\chi}$. Those four channels are XDM $e^{\pm}$, XDM $\mu^{\pm}$, XDM 2-step cascade to $e^{\pm}$ and XDM 2-step cascade to $\mu^{\pm}$. The values for $M_{\chi}$ coincide with the values of $M_{\chi_{2}}$ used in Fig.~\ref{fig:MXDM_1step_2step}, so give good fits to the Fermi data. 

As one can see from Fig.~\ref{fig:ratio_of_synch_rad_4channels}a (B-field I) the power laws describing $J(\textrm{v})$ until $\sim 100$ GHz, vary between:
\begin{equation} 
0.4 \, \le \, \alpha \, \le \, 0.46  \; .
\label{eq:alpha_region_for_1species}
\end{equation}
The channel XDM $\pi^{\pm}$\cite{DarkDiskpaper} is not shown in Fig.~\ref{fig:ratio_of_synch_rad_4channels}, but gives $\alpha = 0.43$ (for $M_{\chi} = 3.1$ TeV). Thus any channel that can be thought (with respect to the injection spectrum of the produced $e^{\pm}$) as a combination of XDM $e^{\pm}$, $\mu^{\pm}$ and  $\pi^{\pm}$ will have a power law in the region of~(\ref{eq:alpha_region_for_1species}), provided $\chi$ can fit the Fermi data and not violate the HESS upper limits\footnote{$\tau^{\pm}$ channels can fit the local $e^{\pm}$ spectra, but are constrained from diffuse $\gamma$-rays, and we do not consider those modes.}. 

The region of~(\ref{eq:alpha_region_for_1species}) is quite narrow\footnote{That is expected from the fact that $\alpha \sim (p-1)/2$ with $p$ the index to the spectrum of the electrons responsible for the radiation (see~(\ref{eq:synch_emiss_power_law_from_elec_spect_power_law}))}, making it a challenge to determine the mass and the annihilation channel based on the microwave Haze data only. Yet this information can be used in understanding the nature of DM. To do this, we will require an accurate measurement of the microwave Haze at frequencies $\sim 100$ GHz, which is challenging due to the various uncertainties involved in subtracting the astrophysical backgrounds and the CMB from the WMAP and Planck data.

\subsection{Dependence of the B-field vs ISRF assumptions}\label{sec:B-filed}

From the Eq.~\ref{eq:synch_emiss} we know that the synchrotron emissivity $j(\textrm{v})$ of an electron inside a magnetic field is proportional to the amplitude of the B-field. If the microwave Haze can be explained from DM, then the region of the Galaxy that provides the major contribution to it is $1 < \, r \, < 2$ kpc. For that region there are significant uncertainties in the literature about the value of the B-fields amplitude and shape(see \cite{Jansson:2009ip,Sun:2007mx,Brown:2007qv,1983ApJ...265..722S,1994A&A...288..759H,Beck:2000dc,Han:2001vf} and \cite{Tinyakov:2001ir} among others). 

To study the effects of the uncertainties of the B-field amplitude, we follow \cite{Cholis:2008wq} and use variants of the default magnetic field profile from GALPROP:
\begin{equation} 
B(\rho,z)=B_{0}exp\left(\frac{R_{\odot}-\rho}{\rho_{c}}\right)exp\left(-\frac{z}{z_{c}}\right).
\label{eq:B-field}
\end{equation}
with $B_{0}=5 \mu G$ being the value for the total(combined random and large scale ``ordered'') local value of the magnetic field, $R_{\odot}=8.5$kpc, $\rho_{c}$ and $z_{c}$ being the characteristic scales along $\rho$ and $z$ directions respectively. 
We used three options for the values of $\rho_{c}$ and $z_{c}$.
\begin{enumerate}
\item{$\rho_{c}=4.5$kpc and $z_{c}=2.0$kpc, defined as ``B-field I'',}
\item{$\rho_{c}=10.0$kpc and $z_{c}=2.0$kpc, defined as ``B-field II''and}
\item{$\rho_{c}=10.0$kpc and $z_{c}=4.0$kpc, defined as ``B-field III''.}
\end{enumerate}
The magnetic field model defined as ``B-field I'', provides agreement with the 408 MHz Haslam map\cite{1982A&AS...47....1H,HazeDMpaper}, while the model defined as ``B-field II'', is the ``conventional'' galprop assumption. ``B-field III'' was chosen to take into consideration uncertainties in the scale height $z_{c}$ of the magnetic field\footnote{Values of $z_{c}\simeq 1$kpc have been used by \cite{Jansson:2009ip} and \cite{Sun:2007mx}. Such a low scale hight for the magnetic field would result in a greater spread of power-laws as those presented in Fig.~\ref{fig:ratio_of_synch_rad} and \ref{fig:ratio_of_synch_rad}, and thus would make it easier to separate among different scenarios.}. In generating the plots shown in Fig.\ref{fig:HAZE_high_freq}, we used the ``B-field I'' model.

In Fig.~\ref{fig:ratio_of_synch_rad}b and c and in~Fig.\ref{fig:ratio_of_synch_rad_4channels}b and c, we show the dependence of the magnetic field on the ratio of synchrotron radiation for the cases studied in Fig.~\ref{fig:ratio_of_synch_rad}a and~\ref{fig:ratio_of_synch_rad_4channels}a.  
We can see that for the cases where $\rho_{c}=10$kpc, the power-law spread is greater than that of $\rho_{c}=4.5$kpc. In the case of  $\rho_{c}=10$kpc, the magnetic field at $1 < z < 2$ (kpc) is weaker by a factor of 2, compared to that of the $\rho_{c}=4.5$kpc case, thus the power in synchrotron radiation is by a factor of 4 lower. As explained previously, one reason for the difference in the observed power-laws is the fact that ICS losses compete with synchrotron losses.  Lower synchrotron radiation losses result in ICS losses being more important\footnote{$\frac{\frac{dE}{dt}_{ICS}}{\frac{dE}{dt}_{total}}$ is greater}. Thus the difference in the observed power-laws shown in~\ref{fig:ratio_of_synch_rad}b and c(~\ref{fig:ratio_of_synch_rad_4channels}b and c) vs~\ref{fig:ratio_of_synch_rad}a (~\ref{fig:ratio_of_synch_rad_4channels}), is enhanced.

Moreover, the weaker magnetic fields shown in \ref{fig:ratio_of_synch_rad}b and c(~\ref{fig:ratio_of_synch_rad_4channels}b and c) versus that shown in~\ref{fig:ratio_of_synch_rad}a (~\ref{fig:ratio_of_synch_rad_4channels}a) results in a lower power-law values for $\alpha$. This simply arises because of the lower total synchrotron energy losses (as we keep our ISRF constant) resulting in a smaller depletion of the higher energy electrons. 

As we see the current uncertainties in the B-field are significant enough to make an analysis based solely on the observed power-laws $\alpha$ degenerate among the various DM models. The ISRF and as well the diffusion properties of the electrons are also not well modeled yet, adding additional uncertainties. Yet, as diffusion models improve and are better constrained, and as more accurate measurements of the ISRF and the B-field in the regions of interest appear, these uncertainties will diminish. Furthermore, one would hope that in these scenarios, we may have some estimate of the(some) DM species from direct searches, or other indirect data. Thus, in concert with such results, such an analysis as we have described would be a complementary tool for understanding the Dark sector.
 
\vskip 0.2in

\section{3 masses}\label{sec:3_masses}
Of course, once two masses are present, we can consider more, and even heavier DM species. We will confine our discussion to the case where there are three DM species and refer to species $\chi_{1}$ as the lightest, $\chi_{2}$ as the intermediate mass species and $\chi_{3}$ as the heaviest species.
\begin{equation} 
M_{\chi_{1}} \, < \,M_{\chi_{2}} \, < \, M_{\chi_{3}}  \; .
\label{eq:3masses}
\end{equation}  
As explained in section~\ref{sec:annih_rates} the three species are expected to have similar fluxes of $e^{\pm}$ at $E \lsim M_{\chi_{1}}$.

A scenario where the three particles are spaced within $100 \gev$-$3\tev$ is essentially indistinguishable from the two-mass case. A scenario where $M_{\chi_{1}} \sim 10$ GeV would be difficult to be seen from the $e^{\pm}$ flux, as Solar modulation is significant at $E \lsim 10$ GeV, so $e^{\pm}$ fluxes at those energies can't be measured accurately. For instance, \cite{Zurek:2008qg} considered a case of two masses where the lightest at $3-10$ GeV is suggested in order to explain via its elastic scattering to NaI target the annual modulation signal measured DAMA/LIBRA \cite{Bernabei:2008yi}, but such a scenario would be difficult to extract from the signals we have discussed. We do note, however, that such a light WIMP could, in principle, produce an additional component of electrons that could contribute to the low energy Fermi Haze, although additional astrophysical sources may contribute as well.

The most interesting from the point of view of locally measured $e^{\pm}$ fluxes is the case where:
\begin{equation} 
M_{\chi_{1}}\sim 100 \textrm{GeV} \, , \,M_{\chi_{2}} \sim 1 \textrm{TeV and} \, M_{\chi_{3}} \sim 10 \textrm{TeV} \; .
\label{eq:3masses_2}
\end{equation}  
where additional contributions at the highest energy can arise.

\begin{figure}[h]
\centering
\begin{tabular}{cc}
\epsfig{figure=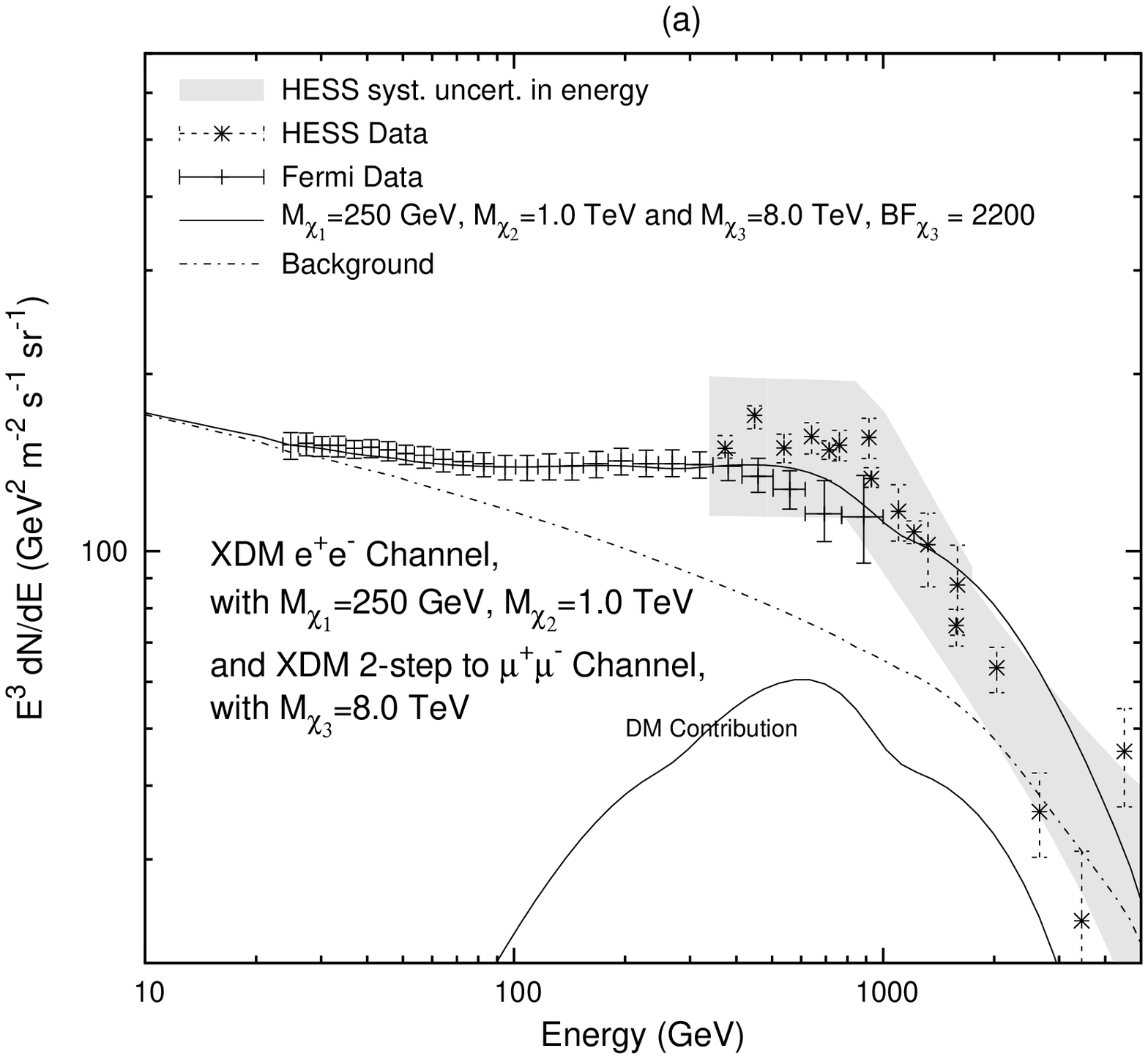,width=3.0in} &
\epsfig{figure=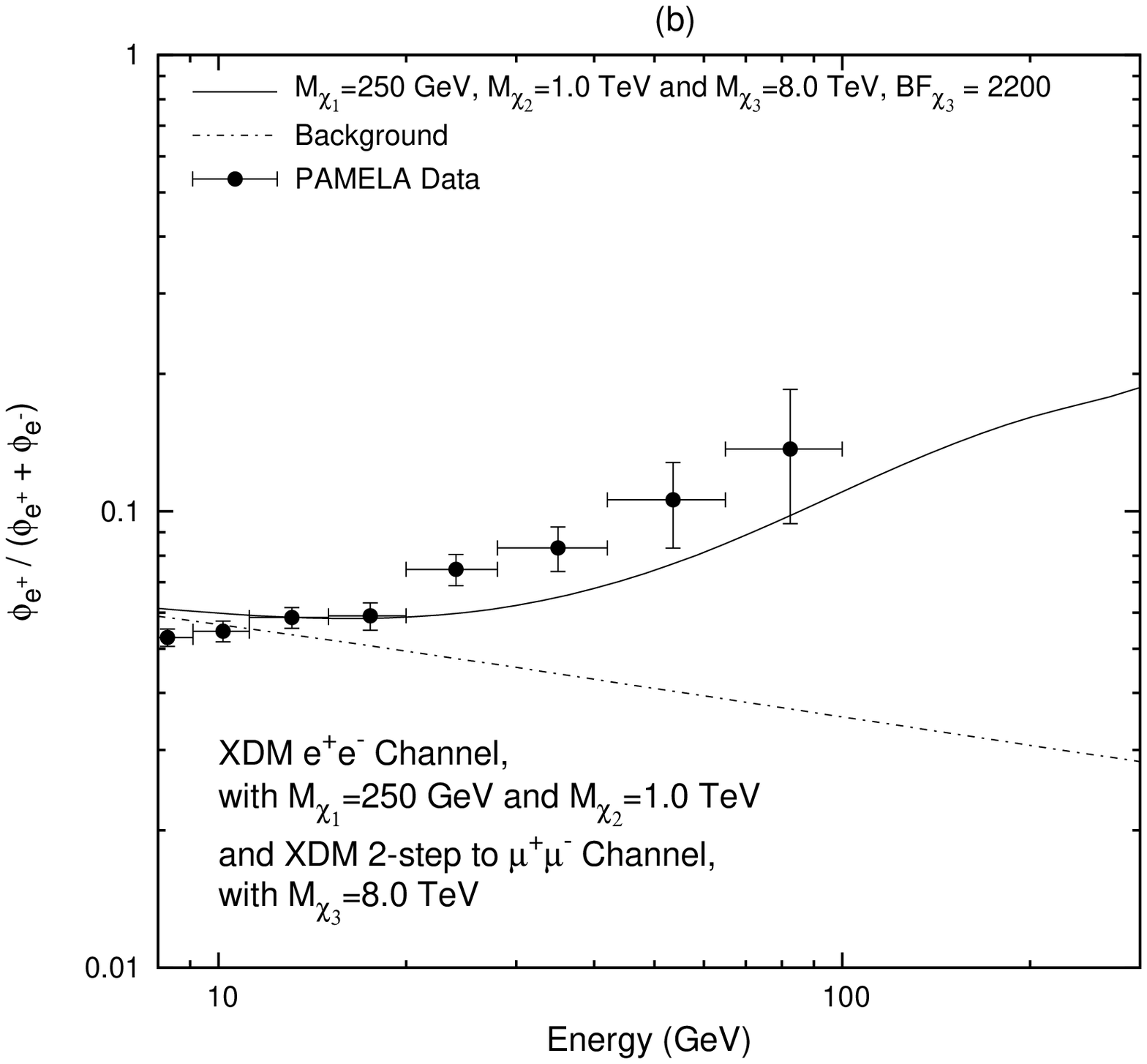,width=3.0in}
\end{tabular}
\caption{Fitted $e^{+}+e^{-}$ spectra as in  Fig. \protect \ref{fig:WIMPspectra} for annihilation of 3 XDM species with $M_{\chi_{1}}=250$ GeV, $M_{\chi_{2}}=1.0$ TeV through $e^{\pm}$ channel and $M_{\chi_{3}}=8.0$ TeV through 2-step cascade to $\mu^{\pm}$.}
\label{fig:MXDM_3masses}
\end{figure}

In Fig.~\ref{fig:MXDM_3masses} we present the case of a $M_{\chi_{1}}=200$ GeV annihilating through XDM $e^{\pm}$ channel, $M_{\chi_{2}}=1.0$ TeV XDM $e^{\pm}$ channel as well and a $M_{\chi_{3}}=8.0$ TeV annihilating through 2-step cascade to muons. We note that even though there is good agreement to the Fermi data, the projected positron fraction is below the one measured by PAMELA by almost a factor of two.

While the two-mass case can be easily realized, the three-mass case described by~(\ref{eq:3masses_2}) is actually allowed only with very specific annihilation modes. The reason behind that is the existence of the HESS $e^{+}+e^{-}$ upper limits\cite{Collaboration:2008aaa,Aharonian:2009ah}, that suggest a turn-off of the source of $e^{\pm}$ responsible for the excess compared to the conventional backgrounds at $\sim 1$ TeV. (A change in the background power law at $\sim 1$TeV would also be a reasonable possibility.) As a $M_{\chi_{3}} \sim 10$ TeV would in general create significant fluxes of $e^{\pm}$ at $\sim 1$ TeV, unless the annihilation mode is adequately soft, there are conflicts with the HESS upper bounds. 

Going to higher than $M_{\chi_{3}}\sim 10$ TeV these tensions are aggravated, as the propagated $e^{\pm}$ contribute to the $e^{\pm}$ flux with $E<5$ TeV. The local primary electron flux at energies $E\gsim 1$ TeV is mainly influenced by local and recent SN explosions\cite{Kobayashi:2003kp,1989ApJ...342..807B,Pohl:1998ug,2001AIPC..587..469B,Strong:2004de,2006AdSpR..37.1909P}, and very young pulsars\cite{1986ARA&A..24..285T,1995PhRvD..52.3265A,Hooper:2008kg,Yuksel:2008rf,Profumo:2008ms,Malyshev:2009tw} as well. These uncertainties make a detection of a heavier mass difficult in the $e^{\pm}$ data.


\section{Scattering and Effects on the DM Halo}\label{sec:scatter}
The presence of ``long-range'' forces in the dark sector can be constrained from the stability of DM halos \cite{Spergel:1999mh,Ackerman:2008gi} and other properties of galaxies \cite{Ackerman:2008gi,Fox:2009in, Feng:2009hw}. However in the context of MiXDM there is an intriguing twist. It is possible that the heavy WIMP could have a lower scattering rate than the lighter WIMP. This may be because of the different mass splittings $\delta$, or the different $m_\phi/m_\chi$, or, possibly, because the WIMPs couple to different force carriers. There are other possibilities, for instance that there are other nearby states, or that there is a residual symmetry allowing degenerate excited states with a long range force between them (analogous to the charged wino, for instance), in which case these light WIMPs might scatter very frequently.  Such scatterings can easily be strong within the dark sector without leading to a significant WIMP-nucleus interaction, as was explored in \cite{Finkbeiner:2009mi}. 

In such a case, if the light WIMP scattering rate is large enough while leaving the heavier WIMP scattering rate small enough, the light WIMP halo would collapse, leading to a completely non-Maxwellian velocity distribution. Since this WIMP is only 1\% of the total energy density, however, this effect would be of little consequence for the total halo. On the other hand, the consequences for direct detection experiments would be profound. The number density and velocity distribution of the lighter WIMPs locally would change significantly, with possibly only the highest velocity particles which are just now falling into the Milky Way present in the event that the light halo has collapsed. In such a case, the lightest WIMP might dominate the annihilation signal from the inner galaxy, while the heavier WIMP would dominate the local cosmic ray signals.

It is not clear how to approach this quantitatively, except to recognize that in MiXDM scenarios the velocity and density structure of the subdominant species can be radically altered, with the subsequent consequences for direct and indirect detection experiments.

\section{Conclusions}\label{sec:Conclusions}

Cosmic ray signals from PAMELA and Fermi among other have prompted a reexamination of our assumptions of the nature of dark matter. Two of the most central are the assumptions that annihilations are into standard model states, and that there is a single parity, namely R-parity, that stabilizes the WIMP.

Relaxing the latter opens the possibility that the dark matter is composed of multiple particles, with different masses and relic abundances. Such as scenario has been previously motivated to explain both DAMA and INTEGRAL. We have argued that such a ``MiXDM'' scenario, in which DM annihilates as in standard XDM into light force carriers can naturally lead to similar annihilation rates, even if the different WIMP relic densities are different by an order of magnitude. The consequence is to modify possible particle physics interpretations of the spectra. Softer spectra can arise naturally from the presence of the second, lighter WIMP state, while even bumps and dramatic spectral features can arise in the positron and electron spectra from the presence of these lighter states.

We find that these models are in good agreement with the Fermi $e^{+}+e^{-}$ flux and the PAMELA positron fraction, with the heaviest mass restricted by the HESS upper bounds. These cross sections are very similar to what is needed to explain microwave and gamma-ray excesses in the inner galaxy region,  while still below bounds from the CMB. Limits from the Galactic Ridge, while possibly constraining, are weakened when velocity-dependent effects are taken into account. 

The presence of the lighter state is possibly detectable in a variety of ways. Beyond the possible features in the $e^+$ and $e^{\pm}$ spectra, the changes in the synchrotron emission spectrum can also help distinguish it. In particular, synchrotron radiation from $e^{\pm}$ of DM origin at frequencies up to $\sim 500$ GHz, probed by Planck, are potentially different for these models. Unfortunately, the uncertainties in the magnetic field, ISRF and electron diffusion, as well as the great number of possible particle physics models, currently it is not possible to discriminate from synchrotron emission alone between a one DM and a two DM species scenarios that both fit the PAMELA and Fermi data. Yet, when combined with a better understanding of all the factors involved in the calculations of the microwave Haze and the synchrotron radiation(B-field, ISRF, diffusion models, dust emission, Planck instrumental errors), such a tool may offer an important handle, especially in combination with other indirect and direct DM searches.

Interesting additional possibilities exist as well. One can consider a third, heavier WIMP, although we find that its effects are highly constrained by the HESS electron limits, and that it is difficult to generate any features at high energies consistent with data. A possibility exists for subdominant components to have very large scattering rates, potentially collapsing the lighter WIMP halo, and radically altering the predictions for direct detection scattering rates.

\vskip 0.2in
\noindent {\bf Acknowledgments}
The authors would like to thank Nima Arkani-Hamed, Gregory Dobler, Douglas Finkbeiner, Joseph David Gelfand, Lisa Goodenough, Ronnie Jansson, Dmitry Malyshev, Tracy Slatyer and Itay Yavin  for fruitful discussions. 
NW is supported by DOE OJI grant \# DE-FG02-06ER41417 and NSF CAREER grant PHY-0449818. IC is supported by DOE OJI grant \# DE-FG02-06ER41417 and also by the Mark Leslie Graduate Assistantship.

\begin{appendix}

\section{DM Halo and CR propagation}\label{sec:DMHalo_CRprop}
\label{sec:CRprop}
For all our calculations we considered an Einasto profile using the parametrization of \cite{Merritt:2005xc}:
\begin{equation}	\rho(r,z)=\rho_{0}exp\left[-\frac{2}{\alpha}\left(\frac{r^{\alpha}-R_{\odot}^{\alpha}}{r_{-2}^{\alpha}}\right)\right],
\label{eq:Einasto_eq}
\end{equation}
with $\alpha = 0.17$, $r_{-2}=25 kpc$ and $R_{\odot}=8.5$ kpc. We considered the local DM mass density to be $\rho_{0}=0.4 GeV cm^{-3}$ that is in agreement with calculations of the local dynamical mass based on Hipparcos data and the estimates on the local baryonic mass\cite{Holmberg:1998xu,Creze:1997rj,Holmberg:2004fj,Korchagin:2003yk} and also with \cite{Catena:2009mf}.

While we assume that our boost from the thermal cross section is purely from particle physics, substructure in the distribution of DM can also exist, affecting the amplitude and spectrum of the signal. For instance, \cite{Hooper:2008kv,Kuhlen:2009is,DarkDiskpaper} find that substructure can create spectral hardening of the propagated $e^{\pm}$ spectrum at high energies. If such substructure gives a significant contribution to the local $e^{\pm}$ flux, it would also enhance the bumps at $E\sim M_{\chi_{1}}$, although this depends on particle physics and astrophysics parameters. 

To determine the cosmic ray properties, we have calculated the propagation of the $e^{\pm}$ in the Galaxy using GALPROP\cite{Strong:2007nh}. In our models for the background $e^{\pm}$ fluxes we have considered a magnetic field of a local value of 5$\mu G$ with a peak value of 33$\mu G$  in the galactic center, and the standard GALPROP Interstellar Radiation Field (ISRF)\cite{Porter:2005qx}. For the diffusion of the CRs galprop considers a diffusion zone within which the spatial diffusion coefficient is considered homogeneous in space depending on rigidity as:
\begin{equation}
D_{xx}=\beta D_{0xx}\left(\frac{R}{D_{rigid \, br}}\right)^{D_{g}},
\label{eq:Diff_eq}
\end{equation}
with $\beta=v/c$, $R=\frac{p}{q}$ the rigidity, $D_{0xx}$ diffusion coefficient at reference rigidity $D_{rigid \, br}$ and $D_{g}$ diffusion index.
Our assumptions for $D_{0xx}$ and $\delta$ provide good agreement with proton, He, C, Fe, B/C and subFe/Fe fluxes, as well as antiproton over proton ratio\cite{DarkDiskpaper}. For a more extensive discussion of background electrons and positrons one can refer to \cite{Strong:2007nh,Cholis:2008vb,Schlickeiser}.

We also define the boost factor that we are going to refer to as just the overall enhancement in the $\sum_{i}n_{i}^{2}\langle\sigma_{i}\mid v \mid\rangle$ needed to fit the data at hand compared to its value assuming an Einasto profile with $\rho_{0}=0.4 GeV cm^{-3}$ and thermally averaged cross sections such that the total relic mass density agrees with the current constraints from WMAP, on $\Omega h^{2}=0.1099 \pm 0.0062$\cite{Hinshaw:2008kr}.

For the case of $M_{\chi_{1}} \ll M_{\chi_{2}}$ and for an Einasto profile as we use in this paper, we can simply write the B.F. as just\footnote{within logarithmic corrections on $M_{\chi_{2}}$}:
\begin{equation}
B.F.\approx \frac{\langle\sigma_{2}\mid v \mid\rangle_{Fitted}}{3\times 10^{-26}cm^{3}s^{-1}},
\label{eq:BF_eq}
\end{equation}
 as $\chi_{1}$ contributes only$\sim \frac{M_{\chi_{1}}^{2}}{M_{\chi_{2}}^{2}}$ to the relic mass density. In our calculations we took into account corrections in the relic density that scale with the logarithm of the masses $M_{\chi}$

\section{Review of Synchrotron Emission}
The emissivity of a single electron to synchrotron radiation is given by\cite{Longair}: 
\begin{equation} 
\label{eq:synch_emiss}
j(\textrm{v})=\frac{\sqrt{3}e^{3}B \sin\alpha}{8 \pi^{2}\epsilon_{0}c m_{e}}\;\frac{\textrm{v}}{\textrm{v}_{c}}\int_{\textrm{v} / \textrm{v}_{c}}^{\infty}K_{5/3}(z)dz
\end{equation}
where $B$ is the magnetic field and $\alpha$ the angle between the B-field and the velocity of the electron, $K_{5/3}$ is the modified Bessel function of the second kind of order $5/3$ and $\textrm{v}_{c}$ is the critical frequency:
\begin{equation} 
\textrm{v}_{c}=\frac{3}{2}\gamma^{2}\frac{eB}{2\pi m_{e}}\sin\alpha.
\label{eq:crit_freq}
\end{equation}
which scales as $\propto \gamma^{2}$, as the gyroradius of the electrons in the ISM is $\propto \gamma$.
The maximum value of $j(\textrm{v})$ is at $\textrm{v}_{peak}=0.286\textrm{v}_{c}$, at $\textrm{v}=3.38\textrm{v}_{c}$ the $j(\textrm{v})$ has $1/10$ of its peak value, while at the very low and very high frequency limits it is given by:
\begin{equation}
\label{eq:synch_emiss_limits}	
j(\textrm{v})=\frac{\sqrt{3}e^{3}B \sin\alpha}{8 \pi^{2}\epsilon_{0}c m_{e}}\;\times\left\{ \begin{array}{ll}
& \frac{4 \pi}{\sqrt{3}\Gamma(1/3)}\left(\frac{\textrm{v}}{2\textrm{v}_{c}}\right)^{1/3} \; \textrm{for} \; \textrm{v} \ll \textrm{v}_{c}\; \\
& \left(\frac{\pi \textrm{v}}{2\textrm{v}_{c}}\right)^{1/2} exp\left(-\frac{\textrm{v}}{\textrm{v}_{c}}\right) \; \textrm{for}\; \textrm{v} \gg \textrm{v}_{c}\;
\end{array}\right\}
\end{equation}

Considering the exponential cut-off at frequencies much higher than the critical frequency we can see that at the highest frequencies to be probed by Planck (up to 857 GHz), the less energetic electrons are not going to emit significant synchrotron radiation. For instance, for a $\simeq$50 GeV electron, and a magnetic field of $\simeq 20\mu$G such as expected in the Haze region, the critical frequency is, $\textrm{v}_{c}\simeq$1 THz, thus its emissivity peaks at 300 GHz. A 1 TeV electron though has for the same B-field a critical frequency of $\textrm{v}_{c}\simeq$400 THz thus the emissivity peaks at $\simeq$120 THz.

\end{appendix}


\bibliographystyle{JHEP}
\bibliography{MXDM}

\end{document}